\documentclass[twocolumn]{aastex631}
\usepackage{appendix}

\newcommand{\mm }{$\mu $m}
\newcommand{\Bra }{Br~${\alpha}$}

\newcommand{\um }{$\mu $m}

\newcommand{\SB}{erg~s$^{-1}$~cm$^{-2}$~arcsec$^{-2}$}
\newcommand{\mJyas}{mJy~$\rm arcsec^{-2}$}

\newcommand{\cmq}{cm{$^{-3}$}}
\newcommand{\cms}{cm{$^{-2}$}}

\newcommand{\kms}{km~s{$^{-1}$}}

\newcommand{\Msol}{M{$_{\odot}$}}
\newcommand{\Mdot}{M{$_{\odot}$~yr$^{-1}$}}

\newcommand{\Lsol}{L{$_{\odot}$}}
\newcommand{\Vlsr}{{V$_{LSR}$}}

\newcommand{\Hii}{H{\sc ii}}

\newcommand{\Ha}{H$\alpha$}

\newcommand{\HII}{H{\sc ii}}

\defcitealias{crowe2024JWST}{Paper I}

\begin{document}
\received{\today}
\revised{\today}
\accepted{\today}

\submitjournal{ApJ}

\shorttitle{Magnetic HII Regions}
\shortauthors{Bally et al.}

\title{The JWST-NIRCam View of Sagittarius C. II. Evidence for Magnetically Dominated HII Regions in the CMZ}


\correspondingauthor{John Bally}
\email{john.bally@colorado.edu}

\author[0000-0001-8135-6612]{John Bally} 

\affiliation{Center for Astrophysics and Space Astronomy, 
     Department of Astrophysical and Planetary Sciences \\
     University of Colorado, Boulder, CO 80389, USA} 


\author[0009-0005-0394-3754]{Samuel Crowe}

\affiliation{Dept. of Astronomy, University of Virginia, Charlottesville, Virginia 22904, USA}

\author[0000-0003-4040-4934]{Rub\'en Fedriani}

\affiliation{Instituto de Astrof\'isica de Andaluc\'ia, CSIC, Glorieta de la Astronom\'ia s/n, E-18008 Granada, Spain}

\author[0000-0001-6431-9633]{Adam Ginsburg}
\affiliation{Department of Astronomy, University of Florida, P.O. Box 112055, Gainesville, FL, USA}

\author[0000-0001-5404-797X]{Rainer Sch\"odel}
\affiliation{Instituto de Astrof\'isica de Andaluc\'ia, CSIC, Glorieta de la Astronom\'ia s/n, E-18008 Granada, Spain}

\author[0000-0002-5306-4089]{Morten Andersen}
\affiliation{European Southern Observatory, Karl-Schwarzschild-Strasse 2, D-85748 Garching bei München, Germany}

\author[0000-0002-3389-9142]{Jonathan C. Tan}
\affiliation{Department of Space, Earth \& Environment, Chalmers University of Technology, 412 93 Gothenburg, Sweden}
\affiliation{Dept. of Astronomy, University of Virginia, Charlottesville, Virginia 22904, USA}

\author[0000-0002-7402-6487]{Zhi-Yun Li}
\affiliation{Dept. of Astronomy, University of Virginia, Charlottesville, Virginia 22904, USA}

\author[0000-0002-6379-7593]{Francisco Nogueras-Lara}
\affiliation{European Southern Observatory, Karl-Schwarzschild-Strasse 2, D-85748 Garching bei München, Germany}

\author[0000-0002-8691-4588]{Yu Cheng}
\affiliation{National Astronomical Observatory of Japan, 2-21-1 Osawa, Mitaka, Tokyo 181-8588, Japan}

\author[0000-0003-1964-970X]{Chi-Yan Law}
\affiliation{Osservatorio Astrofisico di Arcetri, Largo Enrico Fermi, 5, 50125 Firenze FI, Italy}

\author[0000-0002-9279-4041]{Q. Daniel Wang}
\affiliation{Department of Astronomy, University of Massachusetts, Amherst, MA 01003, USA}

\author[0000-0001-7511-0034]{Yichen Zhang}

\affiliation{Dept. of Astronomy, University of Virginia, Charlottesville, Virginia 22904, USA}
\affiliation{Department of Astronomy, Shanghai Jiao Tong University, 800 Dongchuan Rd., Minhang, Shanghai 200240, China}

\author[0000-0002-8389-6695]{Suinan Zhang}
\affiliation{Shanghai Astronomical Observatory, Chinese Academy of Sciences, 80 Nandan Road, Shanghai 200030, People’s Republic of China}


\begin{abstract}
We present JWST-NIRCam narrow-band, 4.05~\mm\ \Bra\ images of the Sgr C \Hii\ region, located in the Central Molecular Zone (CMZ) of the Galaxy.  Unlike any \Hii\ region in the Solar vicinity, the Sgr C plasma is dominated by filamentary structure in both \Bra\ and the radio continuum.   Some bright filaments, which form a fractured arc with a radius of about 1.85 pc centered on the Sgr~C star-forming molecular clump, likely trace ionization fronts. The brightest filaments form a `$\pi$-shaped' structure in the center of the \Hii\ region.   Fainter filaments radiate away from the surface of the Sgr~C molecular cloud.  The filaments are emitting optically thin free-free emission, as revealed by spectral index measurements from 1.28 GHz (MeerKAT) to 97 GHz (ALMA).  
But, the negative in-band 1 to 2 GHz spectral index in the MeerKAT data alone reveals the presence of a non-thermal component across the entire Sgr C HII region. We argue that the plasma flow in Sgr~C is controlled by magnetic fields, which confine the plasma to rope-like filaments or sheets. This results in the measured non-thermal component of low-frequency radio emission plasma, as well as a plasma $\beta$ (thermal pressure divided by magnetic pressure) below 1, even in the densest regions.   We speculate that all mature \Hii\ regions in the CMZ,  and galactic nuclei in general, evolve in a magnetically dominated, low plasma $\beta$ regime.
\end{abstract}


\section{Introduction}\label{sec:introduction}

The Central Molecular Zone (CMZ) surrounding the Galactic Center (GC) of our Milky Way galaxy is a region characterized by extreme conditions.   The densities, temperatures, pressures, turbulent motions, and magnetic field strengths in CMZ molecular clouds are orders of magnitude larger than in the Galactic disk \citep[see, e.g.,][]{Bally87,Bally88,giveon02,ferriere09,kruijssen14,ginsburg16}. However, the current star formation rate in the CMZ \citep[$\sim0.08\,\mathrm{M_{\odot}}$/yr;][]{henshaw23} is around an order of magnitude lower than expected from its 2 to 5$\rm \times 10^7$~\Msol\ reservoir of dense molecular gas \citep{ferriere07}.
Three quarters of the molecular gas is at positive galactic longitudes and radial velocities; most of the compact 24~\um\ sources are at negative galactic longitudes   \citep{Bally88,yusef-zadeh09}. \citet{sormani18} have theoretically explained this asymmetry as the result of unsteady flow of gas in a barred potential.

\begin{figure*}
\centering
\includegraphics[width=1.0\textwidth]{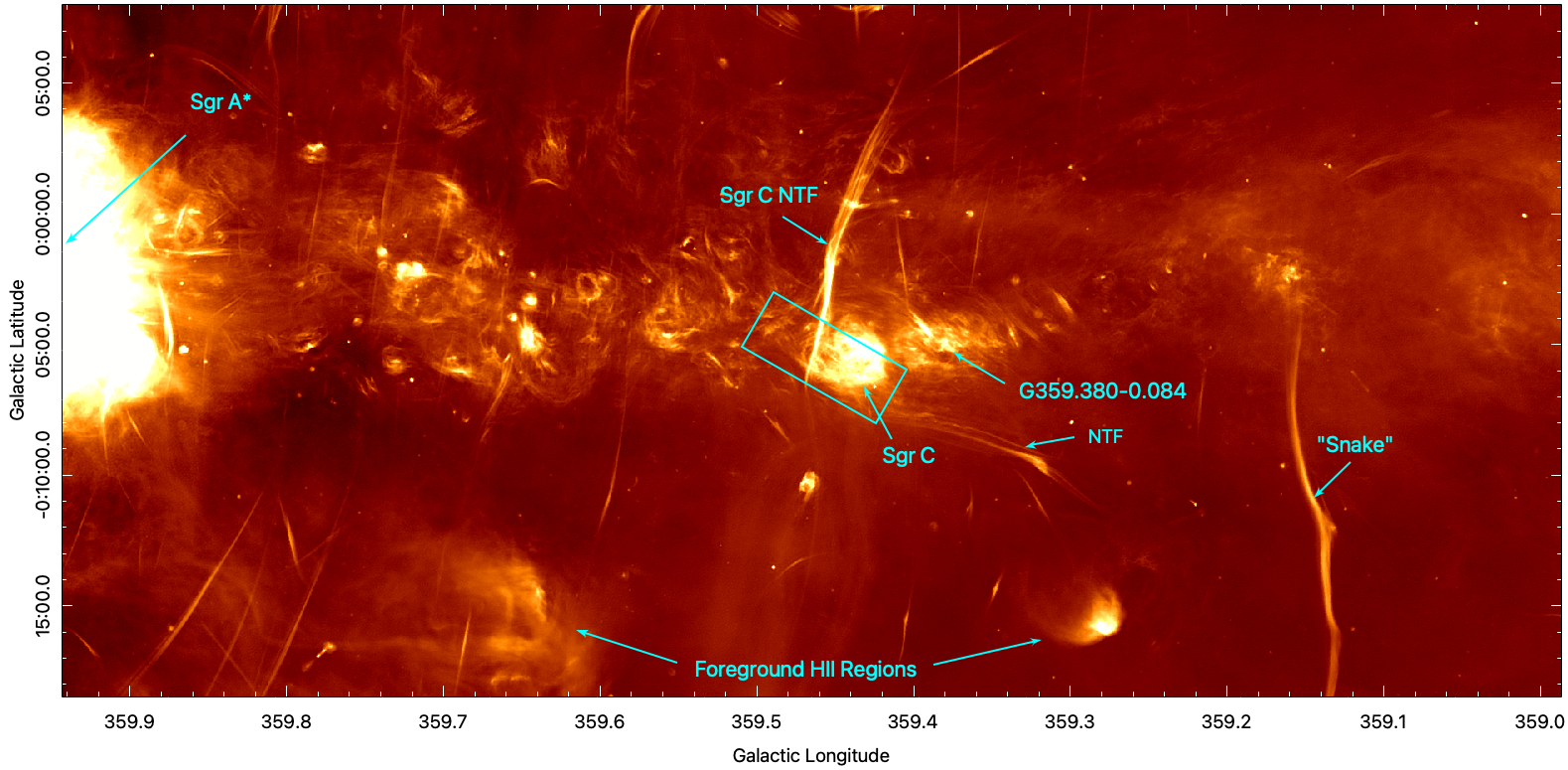}
\caption{\label{fig:widefield_meerkat} The JWST field (cyan rectangle) shown on a MeerKAT radio continuum image centered at 1.28 GHz \citep{heywood22}.   While most NTFs, such as the Sgr C NTF, run orthogonal to Galactic plane, the NTF located to the lower-right of Sgr C is nearly parallel to the Galactic plane, as mentioned in the text.  }
\end{figure*}

The CMZ is strongly magnetized \citep{ferriere09,crocker10,Nishiyama2010,Mangilli2019,Butterfield2024a,Butterfield2024b,pare24}.  Polarized radio continuum emission from hundreds of non-thermal filaments (NTFs) in the CMZ  trace relativistic, magnetically confined synchrotron emission regions \citep{yusef-zadeh84,Yusef-Zadeh2022a,Yusef-Zadeh2022b,Yusef-Zadeh2022c,Yusef-Zadeh2023}.  High-resolution radio observations at 1.28 GHz with MeerKAT \citep{heywood22} demonstrate that filaments pervade the CMZ at a wide range of length scales, from less than 1 pc to over 100 pc. Most NTFs trace either magnetic ropes or edge-on sheets oriented roughly orthogonal to the Galactic plane, thus tracing a poloidal component of the field.   Many NTFs are associated with CMZ star-forming regions containing populations of O-type and Wolf-Rayet stars \citep{yusef-zadeh03}.   
Dust polarization measurements show that the CMZ molecular clouds are threaded by magnetic fields that, on large scales, tend to be roughly parallel to the Galactic plane, thus tracing a toroidal component of the field \citep{Nishiyama2010,Mangilli2019,Guan2021,Hu2022}.  Equipartition arguments applied to the NTFs imply field strengths of about 0.1 to 0.4 mG in the intercloud medium of the CMZ \citep{crocker10,yusef-zadeh22}.  The \citet{chandrasekhar53} method applied to the dust polarization data from CMZ clouds implies 0.1 to 10 mG fields \citep{pillai15,Pan2024}.   
\citet{ferriere09} used the apparent resistance of the NTFs to bending by collisions with 
CMZ molecular clouds to derive field strengths of order 0.1 to 10 mG. \citet{Tress2024} present numerical simulations of the inner Galaxy and show that strong magnetic fields are essential for driving mass inflow from the outer Galaxy onto the CMZ along the leading-edge dust lanes of the Milky Way's stellar bar. 

Sagittarius C (hereafter Sgr C) is the most luminous star-forming region in the negative galactic longitude portion of CMZ \citep{kendrew13}.  Assuming Sgr~C is at the same distance as Sgr  A*, we adopt a galactocentric distance of $8.15\pm0.15$ kpc to the region \citep{reid19}.   Sgr~C has a luminosity $\rm L \sim 10^6$~\Lsol\ and is associated with a prominent \Hii\ region detected from mid-infrared to radio wavelengths \citep{liszt1995,lang10,hankins20}.   \citet{nogueras-lara24} found evidence for $>10^5 ~ \mathrm{M_{\odot}}$ of young stars around Sgr C, with ages of up to $\sim20$ Myr, mixed with a larger, older population with ages over 2 Gyr.  \citet{clark21} classified two bright infrared sources (2MASS J17443734-2927557 and J17444083-2926550) as Wolf-Rayet stars within the Sgr C \Hii\ region using spectroscopy, implying that the OB star population in the Sgr C \Hii\ region has an age of at least $\geq3 ~ \mathrm{Myr}$. Furthermore, Sgr C has been suggested to be a connection point to a stream of gas and dust linking the CMZ and Nuclear Stellar Disk to the Galactic bar \citep{molinari11,henshaw23}, motivating the strong 24 $\mathrm{\mu m}$ emission in this region \citep{carey09} as high amounts of warm dust heated by ionizing stars. This also explains the large gas reserves in this region that are necessary to produce its observed star formation activity  \citep[][\citeauthor{crowe2024JWST}\,\citeyear{crowe2024JWST}; hereafter \citetalias{crowe2024JWST}]{kendrew13, lu20, lu21}.

Spitzer-IRS spectroscopy combined with X-rays and models using the CLOUDY and Starburst99 computer codes indicates that the Sgr C \Hii\ region has an age of $\sim4$ Myr, an electron density $\mathrm{n_e}\simeq300 ~\mathrm{cm^{-3}}$ \citep[which compares reasonably well to the electron density $\mathrm{n_e}\simeq200 ~\mathrm{cm^{-3}}$ derived previously by][]{liszt1995}, and a Lyman photon luminosity of $Q_0 \simeq 1\times10^{50} ~\mathrm{s^{-1}}$ \citep{simpson18a}.  A giant molecular cloud adjacent to the \Hii\ region hosts ongoing massive star formation \citep{kendrew13,lu19b}. \citet{lu24} and \citet{pare24}  show that the cloud has a prominent cometary, head-tail morphology and that the magnetic field orientations follow the cloud edges. However, the dust polarization measurements tend to probe deeper into the cloud rather than the cloud surface.  Nevertheless, the cometary shape and tendency of the field to wrap around the cloud seems to imply an interaction between the adjacent \Hii\ region and the cloud. A similar field morphology is seen in the Messier 16 `Pillars of Creation' \citep{Pattle2018}.  Luminous, massive protostars are found close to the interface between the Sgr C cloud and the Sgr C \Hii\ region \citep[][\citetalias{crowe2024JWST}]{kendrew13, lu20, hankins20}, which either indicates that compressive pressures from the \Hii\ region have triggered star formation or that photo-ablation has uncovered the star-forming region. 

We present JWST-NIRCam observations of the Sgr C \Hii\ region that reveal a uniquely filamentary morphology in the mid-infrared Brackett-$\alpha$ (\Bra ) hydrogen recombination line.  The filamentary structure is also seen in the 3 millimeter (97 GHz) and 23 centimeter (1.28 GHz) wavelength radio continuum.   Inspection of the MeerKAT 1.28 GHz image shows that filamentary morphology is not unique to Sgr C; it is found in all mature CMZ \Hii\ regions, most notably Sgr B1.  None of the foreground \Hii\ regions located in the Galactic disk imaged by MeerKAT exhibit such filamentation, nor is such a morphology seen in other \Hii\ regions in the Galactic disk, such as the Orion Nebula \citep{ODell2001}.  In the Galactic disk, thermal pressure dominates magnetic pressure in the evolution of \Hii\ regions \citep[based on typical Galactic disk field strengths;][]{crutcher12}.   We argue that in the CMZ, the filamentary structure of \Hii\ regions indicates that magnetic pressure is larger than thermal pressure during most of a mature \Hii\ region's evolution.    We present a simple, intuitive description of the expected evolution of magnetically dominated \Hii\ regions.  

\begin{figure*}
\centering
\includegraphics[width=1.0\textwidth]{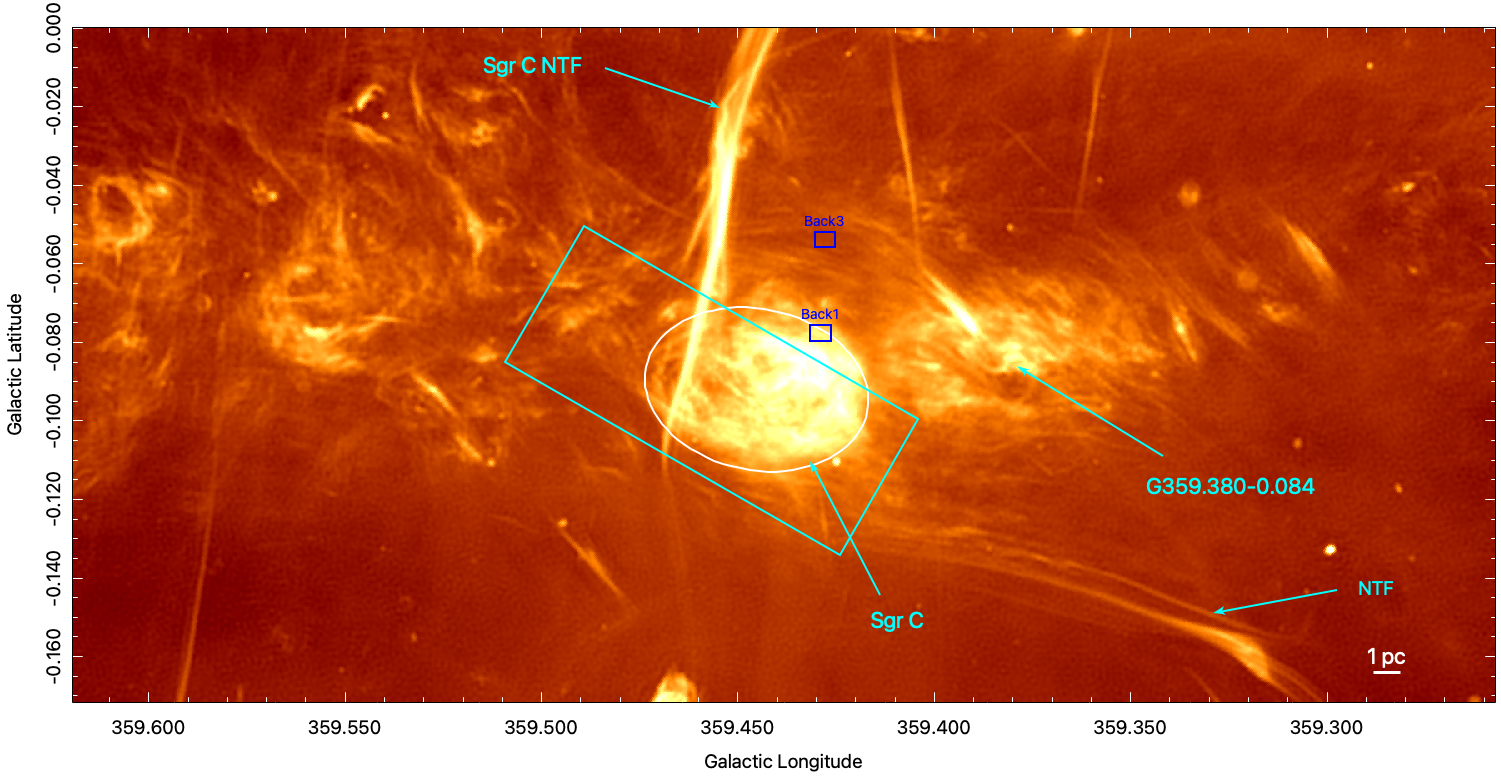}
\includegraphics[width=1.0\textwidth]{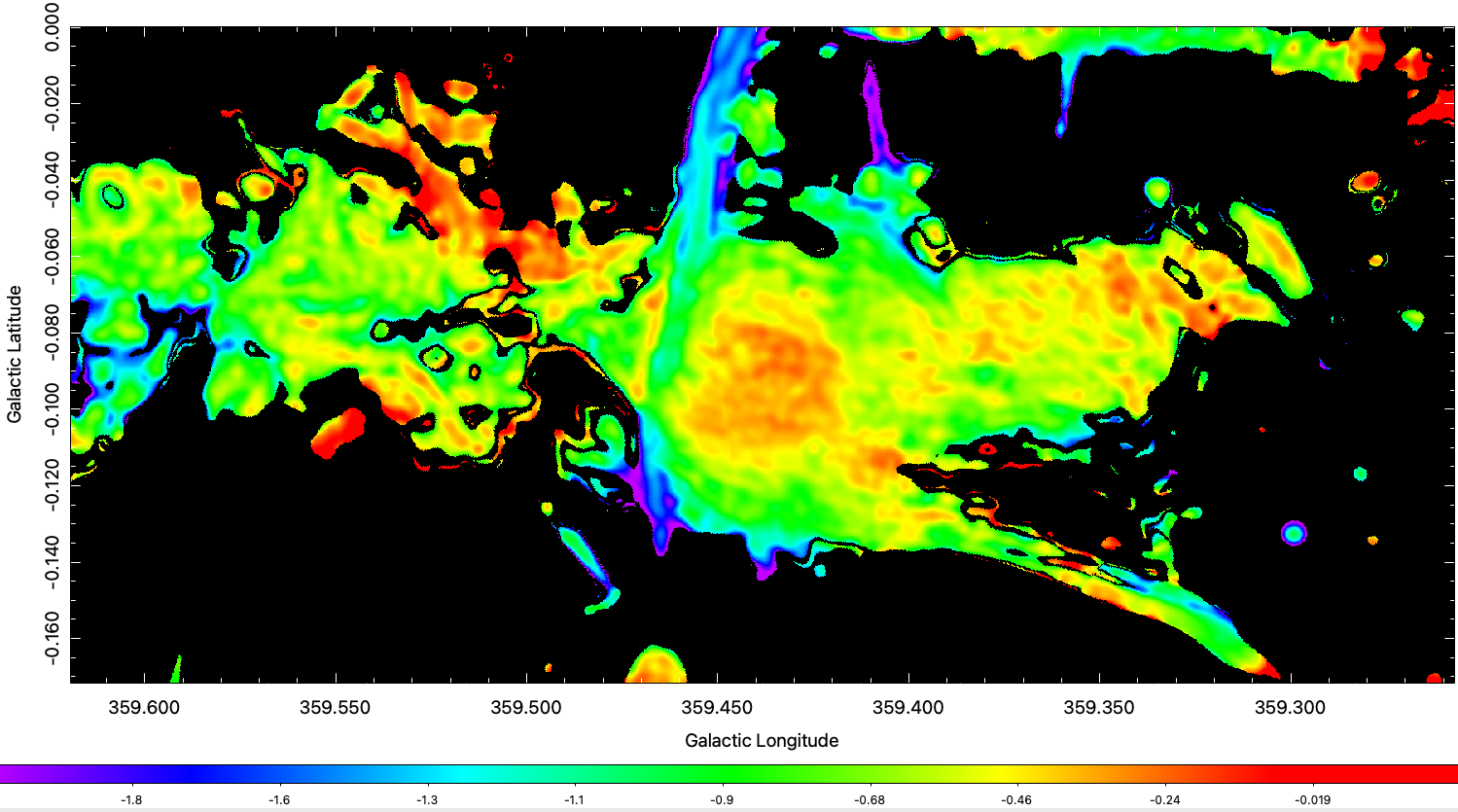}
\caption{\label{fig:mediumfield_meerkat} 
{\bf (Top):} 
Closeup of the  MeerKAT radio continuum image at 1.28 GHz \citep{heywood22} showing the environment of the Sgr~C \Hii\ region.  The cyan boxes labeled Back1 and Back3 are used for background subtraction in the 97 GHz to 1.28 GHz spectral index measurements (\S\ref{tab:spectral_indices}).
Back2 is a smaller box near the core of the \Hii\ region and is shown in Figures \ref{fig:bra_zoomin} and \ref{fig:Pi_filament}.
{\bf (Bottom):} 
MeerKAT spectral index around 1.28 GHz.}
\end{figure*}

\section{Observations}\label{sec:observations}

\subsection{James Webb Space Telescope}\label{sec:jwst_obs}

A single pointing of JWST-NIRCam imaging data was taken of the Sgr C region on the 22$^\mathrm{nd}$ of September 2023 (Program ID: 4147, PI: S. Crowe). The filter set included 8 medium- and narrow-band filters from $\sim1-5$~\um; most relevant to this 
work is F405N, which traces the \Bra\ hydrogen recombination line at 4.05~\um\ and the two adjacent medium-band filters, F360M and F480M. The images taken through the F405N filter were continuum-subtracted by linearly interpolating the stellar fluxes in the medium-band F360M and F480M filters to estimate the mean continuum flux at 4.05~$\mu m$. For further details on the filter set, observational setup, data reduction, and continuum-subtraction of the Sgr C NIRCam data, we defer to \citetalias{crowe2024JWST}.
The final composite image has a field of view of $2.39\arcmin\times5.91\arcmin$  (5.67 $\times$ 14.01 pc).  The long dimension of the image is oriented nearly north-south at a position angle (V3PA) of $1.39^{\circ}$ in Equatorial coordinates and $59.94^{\circ}$ in Galactic coordinates,  causing the pixels to be at an angle with respect to both Celestial and Galactic coordinates.


\subsection{Ancillary Millimeter and Radio Data} \label{sec:ancillary_data}

We use Atacama Large Millimeter Array (ALMA) continuum and spectral line data obtained as part of the ALMA CMZ Exploration Survey (ACES).  Comprehensive details on the ACES survey and data reduction are presented in Ginsburg et al. (in prep.) and Walker et al. (in  prep.).   In the analysis presented here, we use the
3 mm line-free continuum image formed from four passbands used by ACES.  The central frequencies and bandwidths of the four passbands used to form these images are: 
[$\nu$ = 86.2 GHz, $\Delta \nu$ = 0.46 GHz], 
[$\nu$ = 86.9 GHz, $\Delta \nu$ = 0.46 GHz],
[$\nu$ = 98.6 GHz, $\Delta \nu$ = 1.875 GHz],
and 
[$\nu$ = 100.5 GHz, $\Delta \nu$ = 1.875 GHz].    
For a flat-spectrum source, the effective central frequency, weighted by the passbands, is 97 GHz.   We use only the 12-meter ALMA data from ACES, which means that structure on angular scales larger than about 30\arcsec\ is resolved out, while fluxes of features having structure on angular scales smaller than
30\arcsec\ is preserved.   The effective beam size of the ALMA images is $1.97\times1.52\arcsec$.    In addition to the continuum images, we use the ACES data cube featuring the HNCO (4-3) line at 87.925238 GHz. 

We use the total intensity map at 1.28 GHz from the MeerKAT telescope presented in \citet{heywood22} and publicly available in the MeerKAT archive.   This survey includes broad coverage of the entire CMZ region, including higher and lower galactic latitudes than the ACES survey. We also use the MeerKAT spectral index map presented in \citet{heywood22}, which covers the same regions as the total intensity map. The spectral index $\alpha$ is defined by
\begin{equation}\label{eq:spectral_ind_generic}
    \rm F_{\nu} = C\nu^{\alpha}
\end{equation}
where \textit{F} is the flux at a given frequency $\nu$, and
C is a generic constant of proportionality. Therefore, the spectral index $\alpha$ between two frequencies is given by

\begin{equation}\label{eq:spectral_ind_specific}
    \alpha = \frac{\rm log(F_1 / F_2)}{\rm log(\nu_1 / \nu_2)}
\end{equation}. We defer to \citet{heywood22} for further information about the data acquisition, reduction, and availability. 
  
\section{Results}\label{sec:results}

\begin{figure*}
\centering
\includegraphics[width=0.7\textwidth]{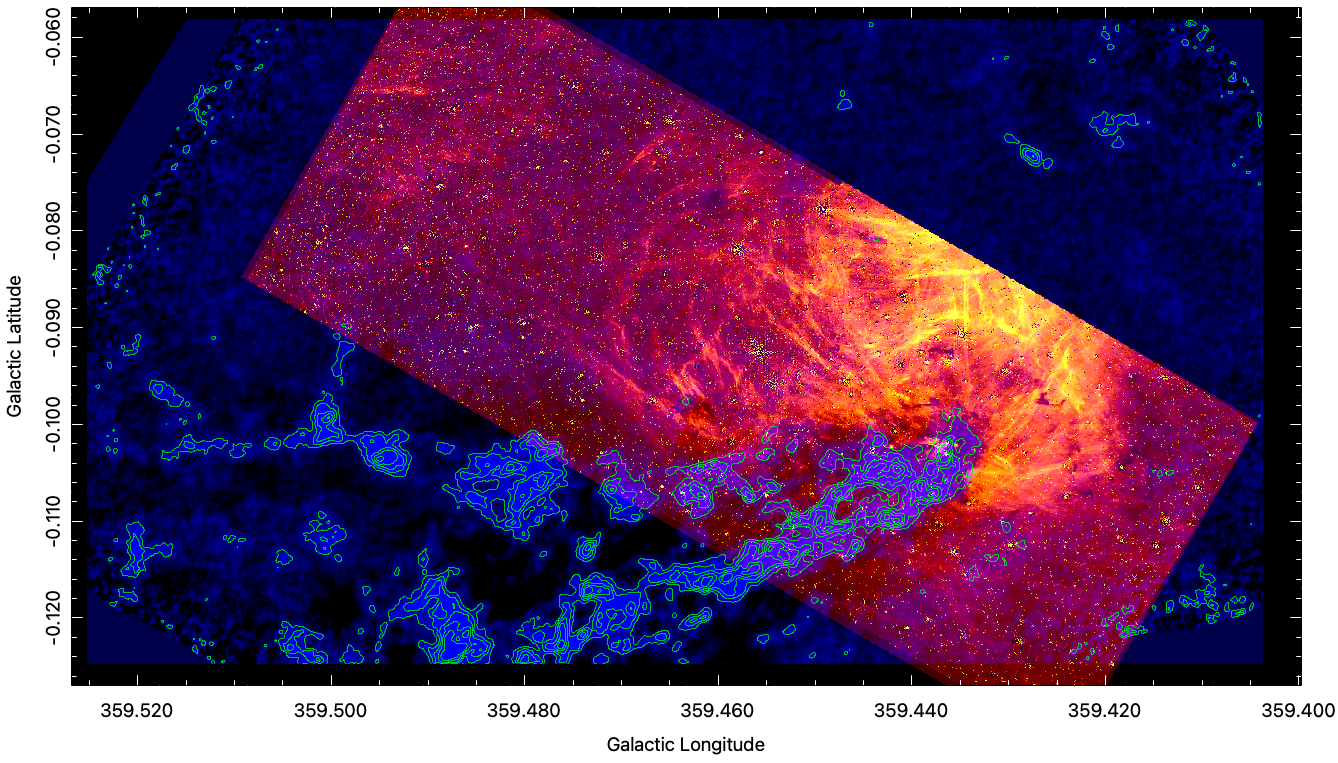}
\includegraphics[width=0.7\textwidth]{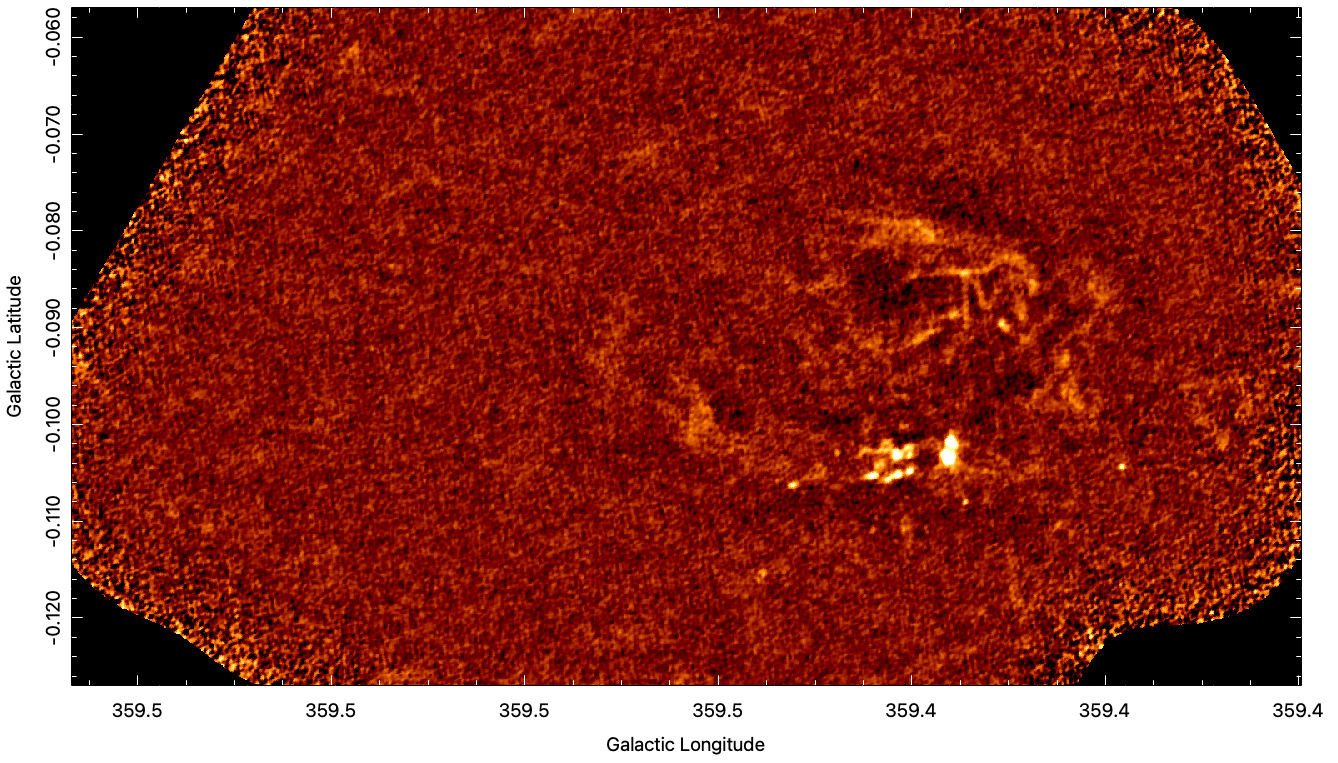}
\includegraphics[width=0.7\textwidth]{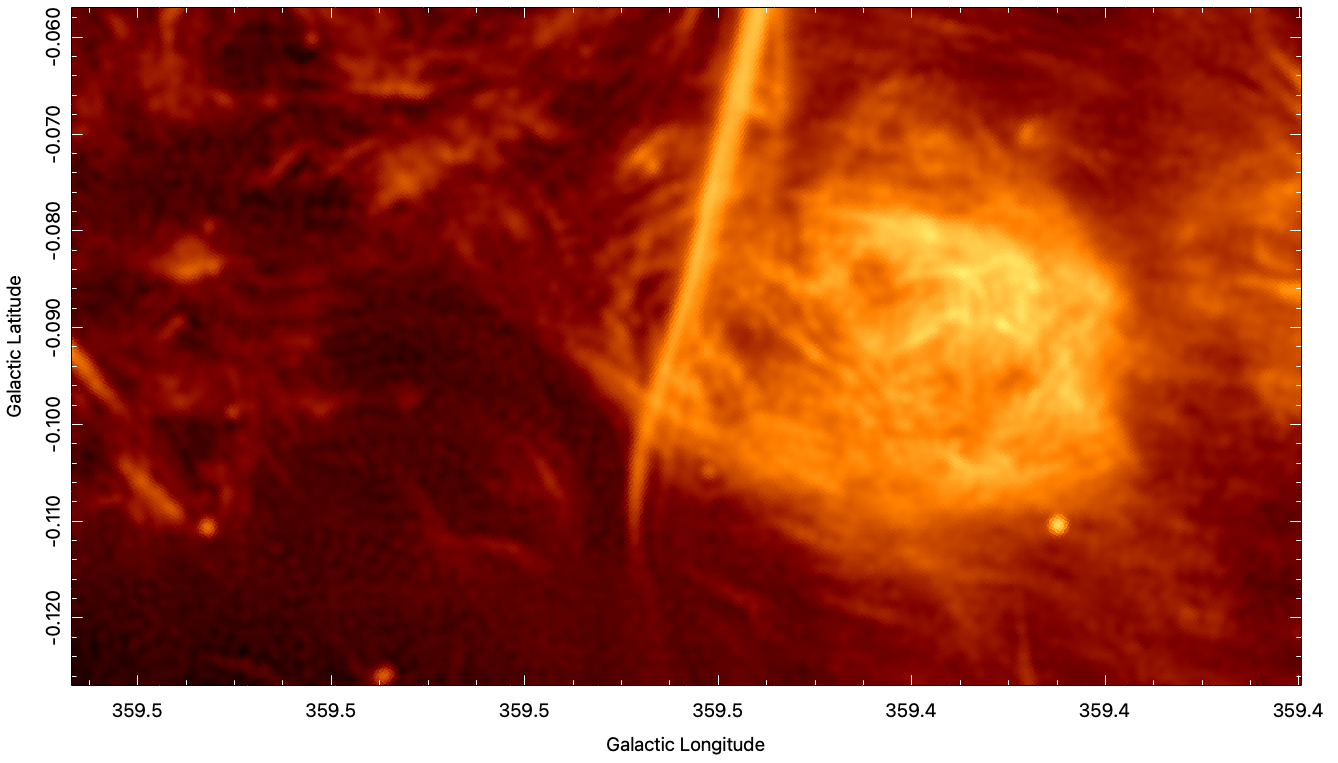}
\caption{\label{jwst_alma_meerkat} 
{\bf (Top):}
\Bra\ emission in the JWST field.  Note the bright `$\pi$-shaped' filaments around [359.434, -0.086]. 
Blue emission and green contours show the intensity of integrated 89.9 GHz transition of HNCO in Jy/beam.  The HNCO map in integtrated over a radial veloctiy range of 41.6 \kms\ centered at \Vlsr\ = -60 \kms .   Contour levels range from 2 to 10 Jy/beam in steps of 2 Jy/beam. The beam FWHM is 1.94\arcsec $\times$ 2.56\arcsec . {\bf (Middle):} ALMA Band 3 97 GHz continuum image with the same FOV as the top panel. {\bf (Bottom):} MeerKAT 1.28 GHz radio continuum image with the same FOV as the above panels.}
\end{figure*}

\begin{figure*}
\centering
\includegraphics[width=1.0\textwidth]{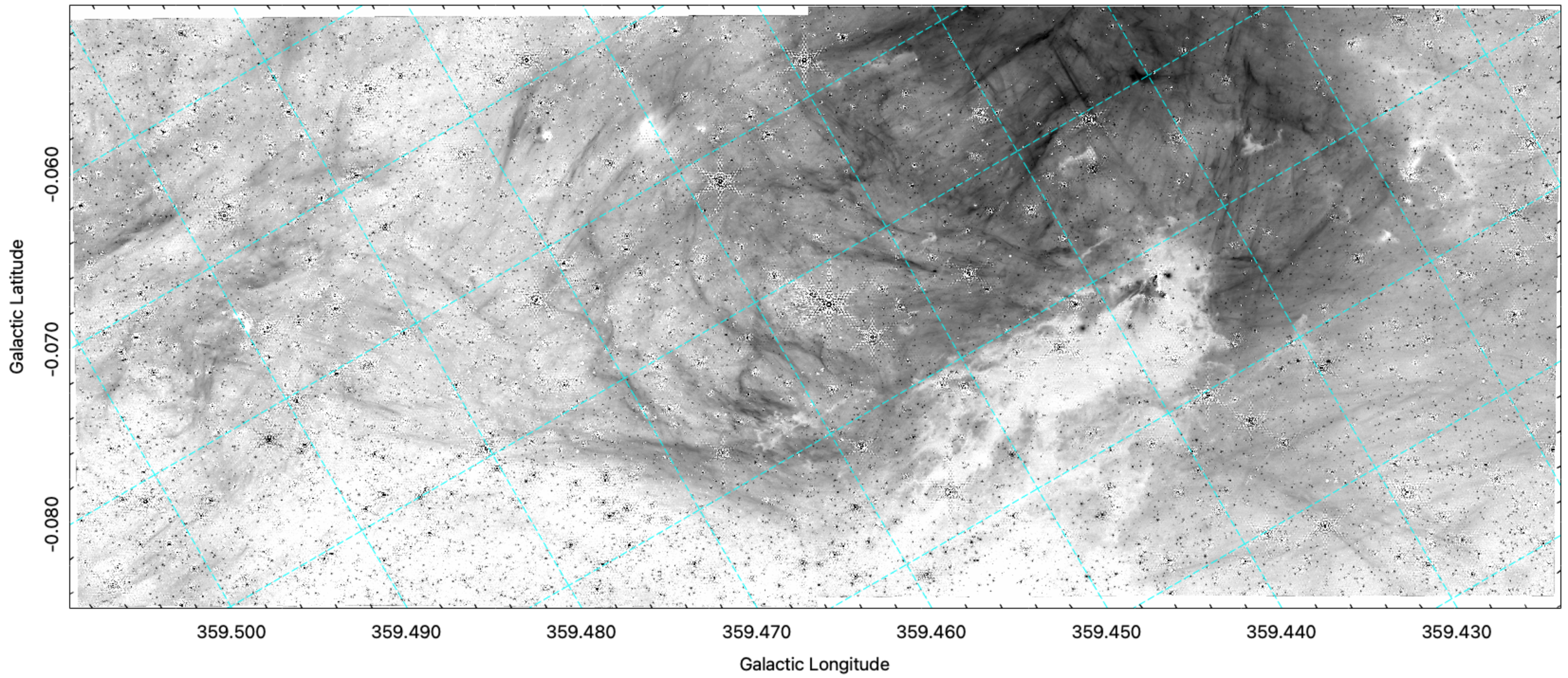}
\includegraphics[width=1.0\textwidth]{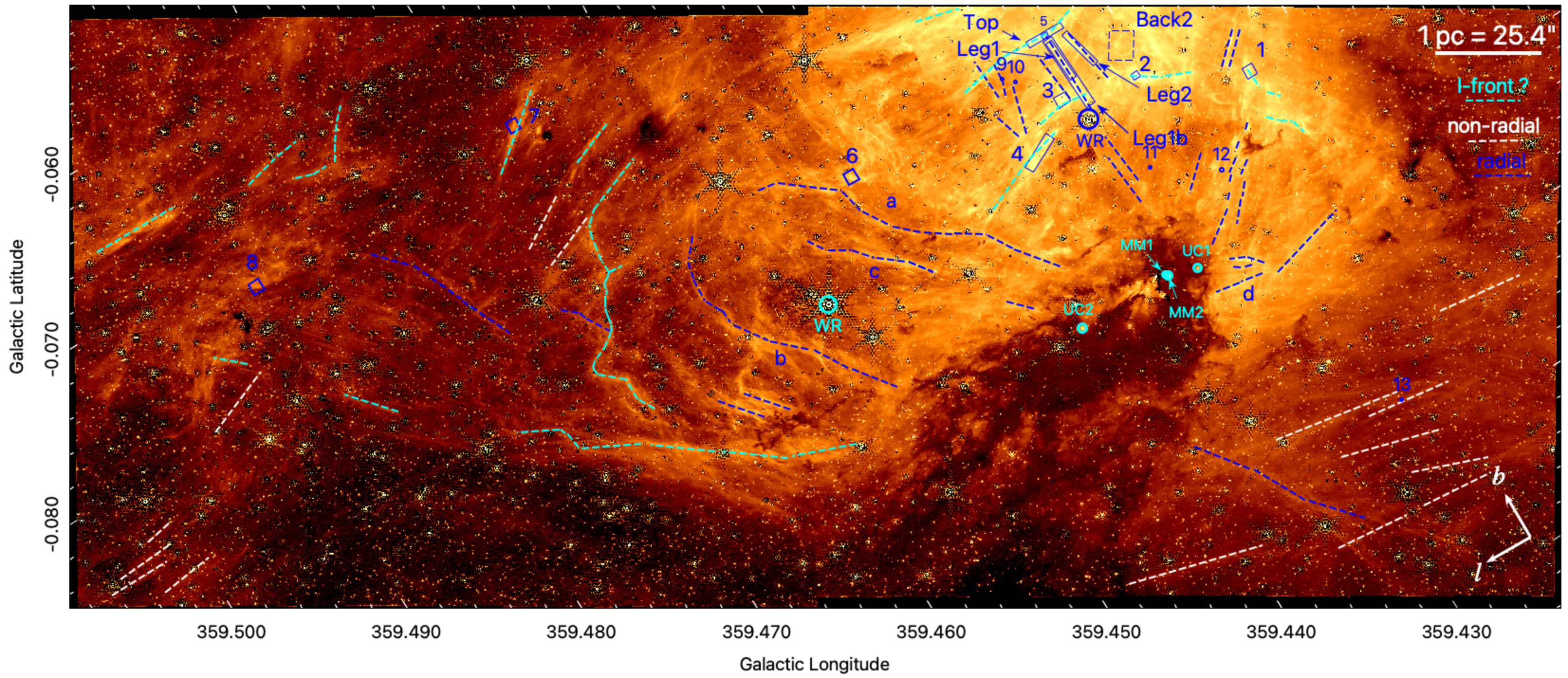}
\caption{\label{fig:bra_zoomin}  
{\bf (Top):}
The JWST \Bra\ image, showing the entire field, optimized to show the fainter filaments away from the \Hii\ region core.  
{\bf (Bottom):}
The \Bra\ image showing some of the various types of filaments discussed in the text, color-coded according to the legend shown in the upper-right corner.  Small, cyan circles show compact \Hii\ regions in the Sgr~C cloud core.  Dashed blue line segments mark filaments mostly pointing away from the Sgr C cloud core.  Cyan filaments trace concentric arcs around the \Hii\ region, and suspected to trace ionization fronts (I-fronts).  Dashed, white line segments mark filaments outside the suspected boundaries of the Sgr~C \Hii\ region.   Blue rectangles and squares show measurement boxes used for radio flux measurements.  The dashed-blue box marks the reference field Back2.  These regions are listed in Table \ref{tab:regions}.   The \Bra\ measurements were made using smaller boxes within some of these regions and are shown in Appendix \ref{section:appendix_magnified_images}. Blue and cyan circles with `WR' underneath mark the locations of the two Wolf-Rayet stars discussed in the text. 
}
\end{figure*}

\subsection{The filaments in Sgr~C}

Figure \ref{fig:widefield_meerkat} shows the 1.284 GHz MeerKAT \citep{heywood22} radio continuum image of a nearly one degree-long portion of the negative longitude section of the CMZ, centered on Sgr C. The cyan box shows the full field of view of the JWST observations presented in this paper. The MeerKAT emission in the upper half of the Figure, most of which originates in the CMZ centered at Galactic latitude -0.050$\arcdeg$, is dominated by clumps and filamentary structure. This includes CMZ \Hii~regions such as Sgr C and Sgr B1 \citep[latter not shown; see Fig. 10 of ][]{heywood22}. Several prominent NTFs such as the `Snake' \citep[see][and references therein]{yusef-zadeh24} near l=359.15$\arcdeg $ and a bright NTF around l=359.45$\arcdeg $ adjacent to Sgr C \citep{yusef-zadeh84} are present. In contrast, there are two foreground \Hii\ regions, located at $\sim$3.5 kpc from the Sun, near the bottom of the figure \citep{Nagayama2009}, which show a relatively smooth morphology.

Figure \ref{fig:mediumfield_meerkat}  shows the environment of the Sgr C \Hii\ region in the MeerKAT 1.28 GHz continuum (top) and spectral index at this frequency (bottom).   Several prominent NTFs are roughly orthogonal to the Galactic plane.  Others, such as the NTF in the lower right (near l=359.32$\arcdeg $), are nearly parallel to the Galactic plane.  The  \Hii\ region is surrounded by a network of filaments that partially wrap around it.  The dominant orientation is at Galactic position angle $\rm GPA \sim$60\arcdeg\ to 75\arcdeg\ (GPA is defined as the 0 towards Galactic north and 90\arcdeg\ along the Galactic plane towards positive Galactic longitudes),  similar to the large-scale field towards the CMZ found by \citet{Mangilli2019}.  
 
The Sgr C \Hii\ region, marked by the large oval in Figure \ref{fig:mediumfield_meerkat},  has an elliptical shape of $\sim$2.5\arcmin\ by 3.4\arcmin\ (5.8 by 8.1 pc) in diameter in the MeerKAT images.  The total flux in this area at 1.28 GHz is $\sim$6.68 Jy and the mean surface brightness is SB(1.28) = 0.278~mJy~arcsec$^{-2}$.
The spectral index in the 1.28 GHz frequency range in the oval is negative, with a value of -0.48 (with an error of $\sim \pm0.05$), indicating non-thermal synchrotron emission.  However, this region includes emission from the bright NTF left from the center of Sgr C.   Excluding this NTF increases the index to about -0.38 to -0.41, with some spots having values as high as -0.2 (with an average error of $\sim \pm0.05$).    The 1.28 GHz morphology and spectral index indicates that there is extended non-thermal emission towards the Sgr C \Hii\ region.  The non-thermal emission may be in-front, behind, or co-mingled with the thermal plasma.  

Figure \ref{jwst_alma_meerkat} shows a continuum-subtracted, 4.05~\mm\ \Bra\ image of Sgr C (top panel), the ALMA 97 GHz continuum image using only the 12-meter antenna data (middle panel), and the MeerKAT 1.28 GHz emission (bottom panel), all with the same FOV.   The \Bra\ emission
from the \Hii\ region is dominated by filamentary structure.  Among the brightest filaments, some are nearly parallel to the Galactic plane, while others are nearly at right-angles, hence roughly orthogonal to the plane, similar to the toroidal and poloidal components of the magnetic field \citep{yusef-zadeh04,Mangilli2019,heywood22}. This phenomenon is examined quantiatively in \S\ref{sec:filfinder}. Additionally, there is a population of fainter filaments that tends to point away from the cometary molecular cloud, seen in silhouette against the \Bra\ emission in the top panel of Figure \ref{jwst_alma_meerkat}.  This panel also shows contours of velocity integrated HNCO emission near 87.9 GHz in steps of 2 Jy/beam taken from the ACES survey. The widths of the filaments range from less than 0.15\arcsec\ (the resolution of JWST at 4.05~$\mu m$) to about 2\arcsec\ (1,000 to 16,000 au)
with the majority being near the resolution limit of JWST.   Most filaments have lengths ranging from 10\arcsec\ to over 100\arcsec\ (0.4 to 4 pc).   

The middle panel in Figure \ref{jwst_alma_meerkat} shows the 97 GHz  continuum emission.   Extended emission on scales larger than about 30\arcsec\ is resolved out.  An elliptical ring of emission traces the outer boundary of the \Hii\ region.   The bright compact sources along the bottom edge trace protostars embedded in the cometary head of the molecular cloud \citep[][\citetalias{crowe2024JWST}]{lu20}. While the (interferometric) ALMA data resolves out most of the extended structure of the Sgr~C \Hii\ region, the most prominent filaments evident in the JWST \Bra\ images are easily visible. 

The bottom panel shows the MeerKAT 1.28 GHz image at the same scale.   Because of the $\sim$4\arcsec\ resolution, some of the filaments seen in the \Bra\ and 97 GHz images are unresolved. However, many are still visible, especially in the extended area around the \Hii~region.

Figure \ref{fig:bra_zoomin} shows \Bra\ emission over the entire field of imaged by JWST in grey-scale (top) and pseudo-color (bottom).  This figure shows that most of the \Bra\ emission originates in an extensive network of filaments.    A sample of filaments are marked by dashed line-segments.   The cyan segments indicate suspected ionization fronts.   These are among the brightest extended \Bra\ features.  Most face the Sgr~C cloud core where four compact but well-resolved \Bra\ features are suspected to trace compact \Hii\ regions (cyan circles marked as UC1, MM1, MM2, and UC2 from right to left) \citep[][\citetalias{crowe2024JWST}]{lu19b}, or one of the two Wolf-Rayet stars marked in the figure.  Suspected ionization fronts tend to have complex, clumpy  morphology with a relatively sharp edges facing the nebular core with a more gradual decline of intensity in the opposite direction.  The blue boxes marked 1 through 5 are examples.   These features trace a segmented arc with a radius of about 1.8 parsecs and centered on the region hosting the suspected compact \Hii\ regions.     Cyan segments and the box labeled 5 appear to trace another ionization front about 2.8 pc from the cloud core.  
The JWST field missed a portion of the \Hii\ region  near the top of the field.  Figure \ref{jwst_alma_meerkat} shows that the upper-edge of the \Hii\ region extends beyond the JWST field to a radius of about 3 pc from the core.  A portion of a suspected ionization front at a radius of $\sim$3.5 pc is marked by a cyan segment just above the label `Top'.   The jagged cyan line near Longitude 359.470 (note that tick marks in the figure are tilted because the JWST field is rotated with respect to Galactic coordinates) may trace a more distant ionization front located about 5.2 pc from the core.  Region 7 and the associated cyan arc traces a $\sim$1 pc-long filament of \Bra\ emission that faces the \Hii\ region that may trace the most distant ionization front from the Sgr~C core so far recognized.  It is located 6.2 pc from the core.  The \Bra\ emission around Legs 1 and 2 and Top form a $\pi$-shaped structure just above the Wolf-Rayet star WR3734.   The mean \Bra\ intensity diminishes with increasing distance up to about 5 or 6 pc from this region.  Comparison of the \Bra\ image with the 1.28 GHz MeerKAT image shows that the `Top' of the $\pi$-shaped filament extends beyond the JWST field towards the upper right.  There is a coherent network of  fainter filaments throughout the JWST field outside the these suspected ionization fronts.  

\begin{figure*}
\centering
\includegraphics[width=1.0\textwidth]{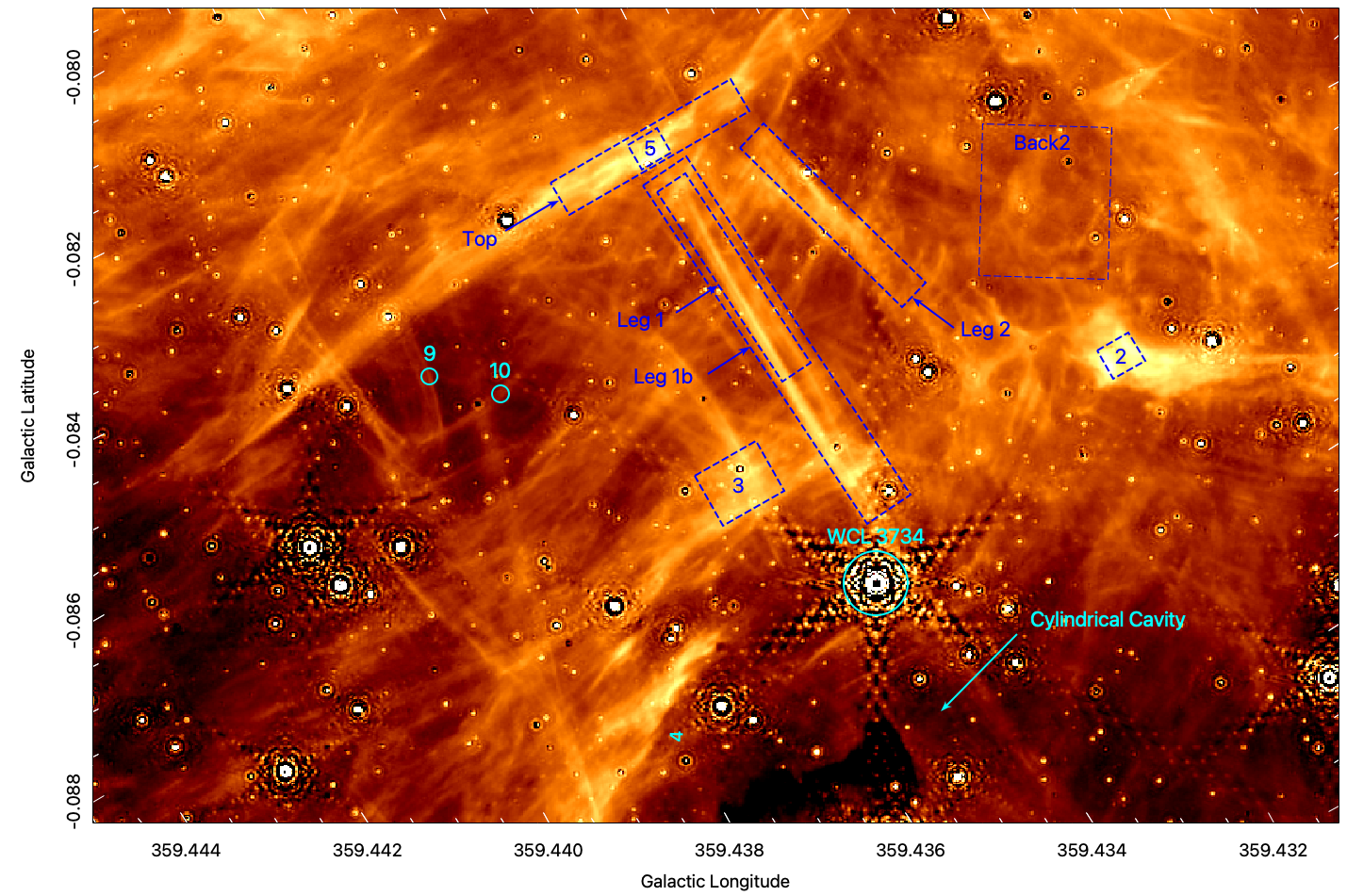}
\caption{\label{fig:Pi_filament} A closeup view of the Br-$\alpha$
emission from the $\pi$-shaped filaments and the cylindrical cavity centered on the Wolf-Rayet star WCL 3734. }
\end{figure*}

Figure \ref{fig:Pi_filament} presents a close-up view of the $\pi$-shaped structure consisting of Leg 1, Leg 2, and Top.   The regions labeled 2, 5, and Top (region 5 is the brightest portion of the `Top' filament) may trace ionization fronts.  However, the feature labeled `Leg 1' has a much smoother appearance and is remarkably straight.  It can be traced for nearly 90\arcsec\ (3.5 pc) from `Top' to nearly the edge of the Sgr C cloud.   It consists of several parallel strands extending over about a 1\arcsec\ to 2\arcsec\ (0.04 to 0.08 pc) wide region.  Each strand has a width of about 0.1\arcsec\ to 0.3\arcsec (0.004 to 0.01 pc).  
A filament about 3\arcsec\ (0.1 pc) left of this bundle runs nearly parallel but is shorter.  This feature crosses a suspected ionization front in region 3.   Region 9 marks the right side of a set of V-shaped filaments ranging in length from 0.2 to 0.4 pc running parallel to Leg 1.   As with Leg1, these filaments also consist of multiple strands barely resolved on the JWST images.   These filaments are nearly orthogonal to the Galactic plane.   A network of fainter filaments runs nearly parallel to the plane and the suspected ionization front marked by `Top'.  Region 10 marks one of these filaments.

These data stand in stark contrast with hydrogen recombination line (\Ha) and radio continuum images of \Hii\ regions near the Sun and in the Galactic disk away from the CMZ, which tend to show mostly smooth plasma morphologies (see e.g. the two foreground \Hii~regions in Fig. \ref{fig:widefield_meerkat}). Although in nearby \Hii\ regions, such as the Orion Nebula, some filaments are seen \citep{yusefzadeh90, ODell2001},  these structures are either ionization fronts, such as the Orion Bar, or are highly tangled, unlike the filaments in the Sgr C \Bra\ image.

\subsection{Filament orientations}\label{sec:filfinder}

To assess the orientation of the \Bra~filaments with respect to the dominant poloidal and toroidal components of the CMZ magnetic field, a statistical analysis was performed using the \texttt{fil\_finder} algorithm \citep{koch15}. This algorithm is typically used for longer wavelength imaging of molecular clouds \citep[e.g., with Herschel FIR and ALMA mm imaging,][]{biswas24,tachihara24,zhou24} or in analysis of cloud simulations \citep[e.g.,][]{mullens24}. Here, we present a novel application of the algorithm for identifying filaments in shorter wavelength ($\lambda \sim 4 \mu m$) hydrogen recombination line emission that could be generalized for use in any region of filamentary ionized gas, in the CMZ or elsewhere.

\begin{figure*}
\centering
\includegraphics[width=0.9\textwidth,angle=180,origin=c]{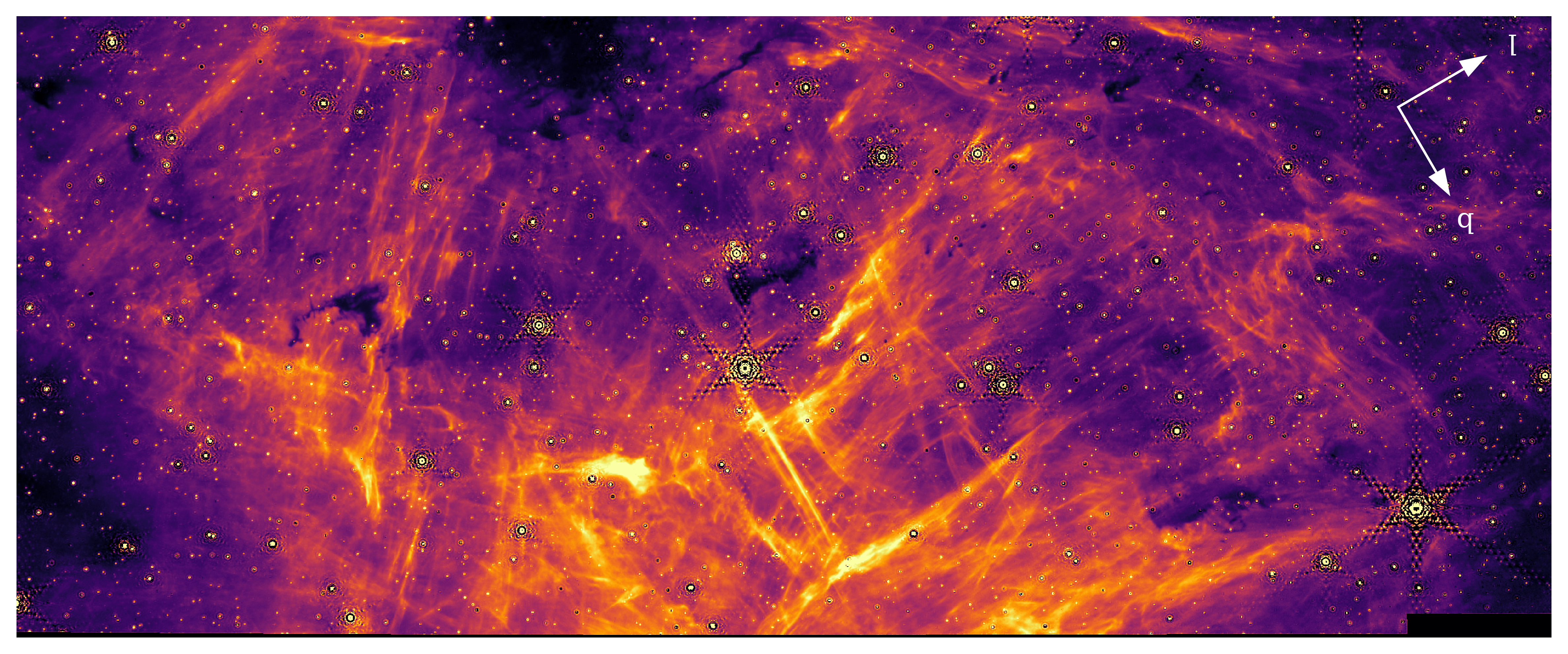}
\includegraphics[width=0.9\textwidth,angle=180,origin=c]{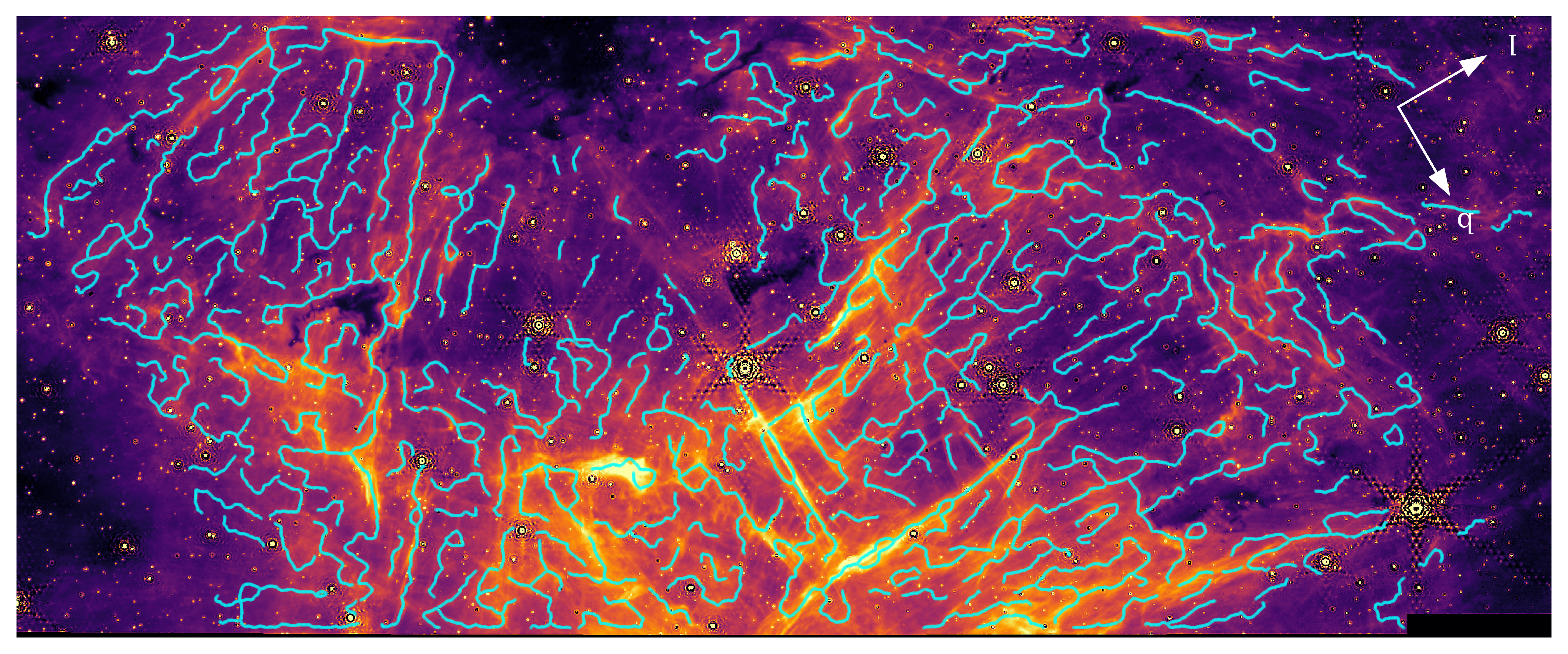}
\includegraphics[width=0.5\textwidth]{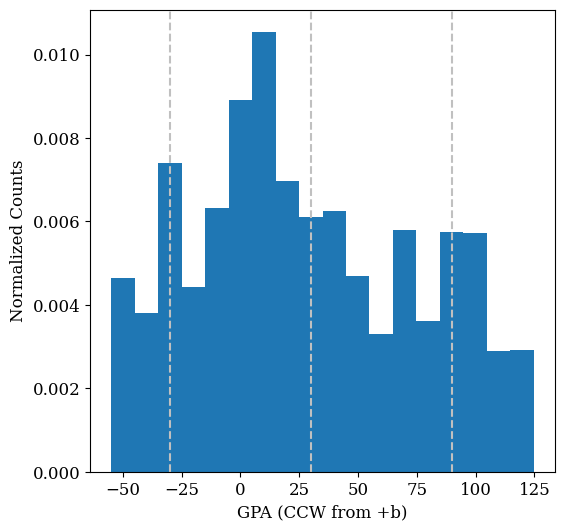}
\caption{\label{fig:filfinder} Top: NIRCam Br$\alpha$ image of the brightest portion of the Sgr C \Hii~region  (see Fig. \ref{fig:bra_zoomin}). Middle: identical image to the top panel with the addition of the filaments and branches identified by the \texttt{fil\_finder} algorithm plotted in cyan. Bottom: Normalized histogram, with $10^{\circ}$ bins, of \Bra~filament Galactic Position Angle (GPA) (defined in the text), weighted by a factor consisting of filament intensity multiplied by filament length.  {\bf Dashed, vertical lines indicate the orientations of the diffraction spikes in the \Bra\:image.}}
\end{figure*}

Following instructions in the \texttt{fil\_finder} tutorial\footnote{https://fil-finder.readthedocs.io/en/latest/tutorial.html}, a mask was first constructed on the brightest portion of the Sgr C \Hii~region with the following parameters: \texttt{glob\_thresh} (i.e., minimum intensity) of 20 MJy/sr, \texttt{size\_thresh} of 1000 pix$^2$, and \texttt{adapt\_thresh} of 0.05 pc. These parameters were chosen to reflect the typical size and brightness of filaments in the \Bra~image, such that most of the visually-identifiable filaments were recovered by the algorithm. We note that several parameter combinations were explored, and that the overall trend of filament orientations (bottom panel of Fig. \ref{fig:filfinder}) was found to be insensitive to different parameter choices.

This mask was then used to construct a `skeleton' consisting of 1-pixel wide representations of each filament. A Rolling Hough Transform \citep[RHT][]{clark14} was performed on each branch of each filament in the skeleton, with a minimum branch length \texttt{min\_branch\_length} of 30 pix., or $\sim10$ resolution elements in the \Bra~image. This returns the orientation of the \Bra~filament.  We note that an orientation could not be measured with the RHT for every filament identified by \texttt{fil\_finder}, likely in cases where, for example, the filament curvature was too high (or its structure too tangled) for the RHT to retrieve a physical result. Therefore, the results presented in Fig. \ref{fig:filfinder} are only for the filaments for which an orientation value was retrieved. The distribution of these filament orientations, weighted by a factor of filament intensity times filament width, is presented in Fig. \ref{fig:filfinder}.

Bright stars exhibit strong diffraction artifacts, mostly produced by the straight edges of the hexagonal elements comprising JWST's 6.5 meter primary mirror.  These show-up as chains of spots in the \Bra~image at Galactic position angle (GPA) of $-$30$\arcdeg$, 30$\arcdeg$, and 90$\arcdeg$. However, because the \Bra\:image has been continuum-subtracted, these artifacts consist of discrete dark and light spots, i.e., regions of positive and negative fluxes, because of the difference in PSF between F405N, F360M, and F480M (the latter two being the `continuum' filters used for subtraction). Therefore, the \texttt{fil\_finder}  algorithm does not appear to detect them (see the middle panel of Fig. \ref{fig:filfinder}).  Nevertheless, the orientation of these artifacts are marked by dashed lines in the bottom panel of Figure \ref{fig:filfinder}.

There appears to be a peak in the filament orientation distribution centered at $\mathrm{GPA} \sim0^{\circ}$, or aligned with Galactic north. This reflects the poloidal component of the CMZ magnetic field (i.e., perpendicular to the galactic plane) \citep{morris07}, which is also observed in the typical orientations of NTFs seen in radio images of the CMZ \citep{yusef-zadeh04,heywood22}.
To quantify the significance of this peak in the distribution of filament orientations, we calculate the `projected Rayleigh statistic' (PRS) \citep[see Eq. 6 and Eq. 7 of ][]{Jow2018} of the distribution.  We find a strongly positive value of $\rm Z_x = 5.8 \pm 0.9$, indicating significant alignment around $\mathrm{GPA} \sim0^{\circ}$, i.e. close to orthogonal to the Galactic plane.   

Besides the peak in filament orientations around $\mathrm{GPA}=0^{\circ}$,  there does not appear to be any other similarly strong preferential orientation of the \Bra\:filaments; i.e., there is no evidence of a multimodal distribution. This is quantitatively supported by the fact that $\rm Z_x>>0$, which indicates strong alignment only near $\mathrm{GPA}=0^{\circ}$. \citep{Jow2018}. This may reflect the preferential  orientation of many filaments with respect to the Sgr C molecular cloud, appearing to either point radially away from the cloud or parallel to the cloud's edge as ionization fronts (see the bottom panel of Figure \ref{fig:bra_zoomin}). This would have the effect of `blurring out' any other subtler systematic orientations due to, e.g., the toroidal component of the CMZ magnetic field. Nonetheless, the preferential alignment of some \Bra\:filaments at $\mathrm{GPA} \sim0^{\circ}$ indicates that there is a population of filaments whose morphology has been strongly influenced by the global CMZ poloidal magnetic field component. However, the large spread of orientations implies that there are other populations of filaments whose orientation is sculpted in larger part by interactions with the Sgr~C molecular cloud, e.g. as ionization fronts.

\subsection{Physical parameters from \Bra }\label{sec:bra_params}

In this section, we derive physical parameters, including the \Bra\ surface brightness, the plasma emission measure, and the electron density of various features such as suspected ionization fronts and filaments in the continuum-subtracted \Bra\ image.  The intensities of most emission lines,  such as the hydrogen recombination lines and the free-free radio continuum emission, are proportional to the emission measure.
The emission measure, EM,  is defined as 
$$
EM = \int n_e^2 dl \simeq n_e^2 L
$$
specified in units of $\rm cm^{-6}~pc$, where $\rm n_e$ is the electron density, dl is a distance element along our line-of-sight (LOS) through the emission region, and L is the effective total path length in parsecs. 

For calculating emission measures, the native JWST pixel values, given in units of MJy~sr$\rm ^{-1}$,  had to be converted to the more conventional surface brightness units,  \SB .  One MJy~sr$\rm ^{-1}$ corresponds to $\rm 2.350443 \times 10^{-5}$~Jy~arcsec$^{-2}$.  The conversion of Janskys to units of surface brightness requires multiplication by bandwidth of the filter in Hertz in order to eliminate the $\rm Hz^{-1}$ in the definition of a Jansky.  The JWST F405N filter has a 0.046 \mm\ passband\footnote{https://jwst-docs.stsci.edu/jwst-near-infrared-camera/nircam-instrumentation/nircam-filters} which corresponds to a bandwidth of 
$\Delta \nu = 8.395 \times 10^{11}$~Hz. 
Thus, 1 MJy~sr$\rm ^{-1}$ corresponds to a surface brightness $\rm 1.971 \times 10^{-16}$~\SB .

Two corrections must be applied to the raw \Bra\ surface brightness estimates before computation of the emission measures and electron densities.  First,  the background emission in the immediate vicinity of a feature, which presumably originates in front or behind the emission region under consideration, must be subtracted.  Second, the fluxes must be corrected for extinction.

Measurement of the fluxes and physical properties of various compact features and filaments require careful background subtraction because of emission along the LOS that is unrelated to the feature under consideration.  
For each feature, elongated measurement boxes (`DS9 region files') are used `on' and `off' the suspected feature.  The `off' positions are taken to be a similar sized region deemed to be free from emission from the feature, determined uniquely for each feature.  The resulting fluxes and derived parameters are therefore sensitive to the choice of `off' region used for background subtraction.   The uncertainty in the measured fluxes is estimated to be about 20\% to 30\%  based on the typical range of background fluxes around each region.    The high density of stars in the CMZ fields is the major source of this uncertainly.  In the continuum subtracted images, bright stars leave large residuals and fainter stars are poorly subtracted due to their color variations.  We list the coordinates of the fields for which measurements were made in Table \ref{tab:regions}, their areas,  and orientations.   The positions used to estimate nearby background fluxes are shown in the figures in Appendix \ref{section:appendix_magnified_images}. Because of the vastly different resolution of the radio and \Bra\ images, different regions are used for the infrared and radio measurements.  

In Table~\ref{tab:regions} the first entry, `\Hii', corresponds to the large oval shown in Figure \ref{fig:mediumfield_meerkat} encompassing the entire Sgr~C \Hii\ region.   The dimensions listed give the semi-major and semi-minor axes of the ellipse. The next four entries (Leg1, Leg1b, Leg 2, and Top) correspond to rectangular apertures shown in Figure \ref{fig:bra_zoomin}, which mark the `$\pi$-shaped' group of filaments located near the top of the JWST \Bra\ image in the brightest portion of the \Hii\ region.   The numbered entries correspond to bright features also marked on Figure \ref{fig:bra_zoomin}. Entry 5 is located inside the rectangle `Top'  delineating the top of the $\pi$-shaped filaments.  Regions 3 to 13 correspond to other features discussed below.  The final three regions Back1, Back2, and Back3, refer to areas where backgrounds were measured on the MeerKAT 1.28 GHz image.  Magnified views of sub-fields in the \Bra\ image are shown in Appendix \ref{section:appendix_magnified_images} along with the locations various regions used for \Bra\ measurements listed in Table~\ref{tab:regions}.    The physical properties of some selected filaments,  suspected ionization fronts, and other features are tabulated in Table \ref{tab:filament_params} and discussed below.  

\citet{nogueras-lara24} measured  the near-infrared reddening to thousands of stars to create a map of the  2.15 \mm~K$_s$ extinction toward the Sgr~C \Hii\ region (see their Figure 3).  \citet{nogueras-lara24} find that the extinction in the K$_s$ filter varies from $\rm A_{Ks}\sim2.0$ to 2.5 magnitudes towards the \Hii\ region, but reaches values larger than 3 magnitudes towards the Sgr~C molecular cloud, which appears nearly opaque in the \Bra\ image.   Around the brightest part of the nebula, containing the  $\pi$-shaped filaments,  $\rm A_{Ks}\sim2.0$ magnitudes.    We scale the 2.15 \mm\ extinction to the 4.05 \mm\ wavelength of the \Bra\ emission line using the relation 
$\rm A_{\lambda} \propto 1/ \lambda $ \citep{fitzpatrick99}.    
Thus, $\rm A_{4.05 \mu m} \approx 0.529 A_{Ks}$.   The extinction corrected surface brightness values, SB, are related to the raw surface brightness measurements, SB$_o$ by  
$$
SB  \approx SB_o  ~10^{0.529 A_{Ks}/2.5}
$$
which implies scaling by a factor f = 2.65 to 4.31 for $\rm A_{Ks}$ = 2 to 3 magnitudes.   Uncertainties in the reddening correction due to incomplete knowledge of the extinction curve in the CMZ and spatial variations introduce uncertainties in the extinction corrected values. If a CMZ extinction law of $\rm A_{\lambda} \propto 1/ \lambda^{1.5} $ was instead adopted, similar to more recent extinction measurements towards the CMZ \citep[see][noting the uncertainties in the extinction curve between $\sim2-4\:\mu m$ due to the presence of several prominent emission features]{nishiyama09,fritz11}, each \Bra\:surface brightness measurement would decrease by $\sim23\%$ (and the corresponding electron density, thermal pressure, and plasma $\beta$ by $\sim12\%$). Thus, the adopted extinction law of $\rm A_{\lambda} \propto 1/ \lambda $ may be a conservative overestimation of the strength of thermal pressure (Eq. \ref{eq:thermal_pressure}) in the Sgr C \Hii~region.

The emission measure is related to the intensity of the visual wavelength 
($\rm \lambda = 6563 $~\AA ) Balmer-alpha (\Ha ) emission line, 
$$
\rm EM = 3.975 \times 10^{17} I(H_{\alpha} )  ~~~~~~~~ \rm (cm^{-6}~pc)
$$
where I(\Ha ) is in \SB\ units \citep{Haffner2003}.  We use the Case B hydrogen recombination spectrum to relate the expected intensity of \Bra\ to \Ha\ at 5,000 and 10,000 K.    Using Table 14.2 in \citep{Draine11}, we find that I(\Ha )$/$I(\Bra ) = 29.6023 at 5,000 Kelvin and 36.0833 at 10,000 Kelvin.  At these two plasma temperatures, the emission measure is given by 
$\rm EM = 1.1766 \times 10^{19}  I(Br_{\alpha} )$  at T=5,000~K  
and
$\rm EM = 1.4343 \times 10^{19}  I(Br_{\alpha} )$  at T=10,000~K.
Interpolating to 6,000~Kelvin \citep[which we take to be the typical \Hii~region temperature in the CMZ;][]{Wink1983,Balser2011} gives
$$ \rm
EM = 1.2282 \times 10^{19}  I(Br_{\alpha} )~~~~~~~~ \rm (cm^{-6}~pc)
$$
where the \Bra\ intensities are the extinction-corrected values.   The mean and peak electron densities are derived by assuming that the LOS depth of a feature, L,  is identical to the projected width on the plane of the sky, corrected for the resolution of JWST.  Thus,

$$ \rm
  n_e \approx \sqrt {EM \over L} ~~~~~~~~(cm ^{-3} ).
$$

This approximation works for cylindrical structures, but for edge-on or inclined sheets or bubbles, it will underestimate L and overestimate $\rm n_e$.

The plasma thermal pressure, $\rm P_T$, defined as 
\begin{equation}\label{eq:thermal_pressure}
    \rm P_{\sc ii} = 2 n_{\sc ii} k T_{\sc ii}
\end{equation}
where the factor of two accounts for the contribution of both protons and electrons to the pressure in terms of the electron density $\rm n_{\sc ii}$, k is the Boltzmann constant, and $T_{\sc ii}$ is the plasma temperature, which we take to be 6,000 Kelvin \citep{Wink1983,Balser2011}.    The plasma $\beta$ parameter is a measure of the importance of magnetic fields for the dynamics of the plasma.  The plasma $\beta$ is defined as
\begin{equation}\label{eq:plasma_beta}
\beta = \rm P_T / P_B
\end{equation}
where $\rm P_T$ is the thermal pressure (Eq. \ref{eq:thermal_pressure}) and $\rm P_B$ is the magnetic pressure, 
\begin{equation}\label{eq:magnetic_pressure}
\rm P_B = \frac{B^2}{8 \pi}
\end{equation}
where B is the magnetic field strength. For the analysis here, we assume that the magnetic field has a strength of 1 mG in every region, the typical large-scale average field strength in dense CMZ clouds \citep{ferriere09, pillai15}.   

\begin{table}
	\centering
	\caption{Regions used for radio and \Bra\ flux measurements. }
	\label{tab:regions}
	\begin{tabular}{lcccc} 
		\hline
		Region	&  Longitude   & Latitude  & Dimensions & GPA   \\
                 & (deg)       & (deg)     & (arcsec)   &  ($^{\circ}$) \\
        \hline
\Hii\   & 359.4450 & -0.0922  & $ 73.82  \times 103.48  $ & 78   \\
Leg 1   & 359.4345 & -0.0863  & $ 1.57   \times  10.41  $ &  3   \\
Leg 1b  & 359.4344 & -0.0872  & $  2.39  \times  18.64  $ &  3   \\
Leg 2   & 359.4330 & -0.0863  & $ 1.57   \times  10.41  $ & 15   \\
Top     & 359.4355 & -0.0843  & $ 9.54   \times   1.71  $ &  0   \\
1       & 359.4239 & -0.0928  & $ 0.130  \times   1.120  $ &  0   \\
1b      & 359.4242 & -0.0928  & $ 0.130  \times   1.120  $ &  0   \\
2       & 359.4309 & -0.0896  & $ 0.384   \times  0.707  $ &  0   \\
2b      & 359.4306 & -0.0896  & $ 0.258   \times  0.693  $ &  0   \\
3       & 359.4357 & -0.0886  & $ 3.23   \times   2.64  $ &  0   \\
4       & 359.4388 & -0.0909  & $ 9.28   \times   2.42  $ & 28   \\
5       & 359.4346 & -0.0844  & $ 1.57   \times   1.31  $ &  0   \\
6       & 359.4504 & -0.0860  & $ 0.94   \times   0.94  $ &  circle   \\
7       & 359.4679 & -0.0719  & $ 1.956   \times   0.226  $ &  328   \\
8       & 359.4878 & -0.0727  & $ 0.527   \times   0753  $ &  328   \\
9       & 359.4385 & -0.0855 & $ -  $ &  peak pixel   \\
10      & 359.4378 & -0.0861 & $ -  $ &  peak pixel   \\
11      & 359.4329 & -0.0954 & $ -  $ &  peak pixel   \\
12      & 359.4289 & -0.0979 & $ -  $ &  peak pixel   \\
13      & 359.4263 & -0.1171 & $ -  $ &  peak pixel   \\
Back 1  & 359.4288 & -0.0779  & $ 18.67  \times  13.99  $ &  0   \\
Back 2  & 359.4305 & -0.0875  & $  6.97  \times   5.94  $ & 58   \\ 
Back 3  & 359.4276 & -0.0541  & $ 18.67  \times  13.99  $ &  0   \\ 
		\hline
	\end{tabular}
\end{table}

Table \ref{tab:filament_params}
lists measurements derived for selected regions. 
The 1st column gives the identity of the region as marked in the Figures.   
Column 2 gives the locally {\it background subtracted}  mean counts in each measurement box in MJy~sr$\rm ^{-1}$. 
Column 3 gives the extinction corrected surface brightness of the \Bra\ emission.
Column 4 gives the assumed LOS path-length, L, taken to be equal to the minor-dimension width of the feature on the plane of the sky.  
Column 5 gives the emission measure.  
Column 6 is the derived electron density $n_e$ for the assumed path length, L.    
Column 7 lists the resulting plasma thermal pressure, $\rm P_T$. 
Column 8 gives the plasma $\beta$ parameter.    
For each region, we evaluate the parameters for two values of extinction, corresponding to A$\rm _{K_s}$ = 2 magnitudes (first row for each region) and A$\rm _{K_s}$ = 3 magnitudes (second row).

\begin{table*}
	\centering
	\caption{Physical parameters of the brightest  features derived from the \Bra\ image.  }
	\label{tab:filament_params}
	\begin{tabular}{lcccccccl} 
		\hline
Region	& counts &    Corrected SB & L &  EM  & $\rm n_e $ & $\rm P_{ii}$ & $\beta$ & Comments \\
  &  (MJy/sr) &   ($\rm erg~s^{-1}cm^{-2}as^{-2}$ )&  (pc) & ($\rm cm^{-6} pc$) & (\cmq )  & (\rm dynes~cm$^{-2}$)  &  & \\
        \hline
Sgr C  (A$_{Ks}$=2)  & 25.1  & 1.53e-14  & 5.83  & 1.88e+05  &   179  & 2.97e-10  & 0.0075 & Entire \Hii\ region \\
Sgr C  (A$_{Ks}$=3)  & 25.1  & 1.95e-14  & 5.83  & 3.06e+05  &   229  & 3.80e-10  & 0.0095 & \\
L1 (A$_{Ks}$=2)   & 50.1  & 2.62e-14  & 0.00966  & 3.22e+05  &  5770  & 9.56e-09  &   0.24 & $\pi$-leg \\
L1 (A$_{Ks}$=3)   & 50.1  & 4.27e-14  & 0.00966  & 5.24e+05  &  7367  & 1.22e-08  &  0.307 & \\
L1 peak           & 83.8  & 4.38e-14  & 0.00966  & 5.37e+05  &  7459  & 1.24e-08  &  0.311 & \\
                  & 83.8  & 7.13e-14  & 0.00966  & 8.76e+05  &  9524  & 1.58e-08  &  0.397 & \\
L2                & 37.3  & 1.95e-14  & 0.00848  & 2.39e+05  &  5312  & 8.8e-09   &  0.221 & $\pi$-leg \\
                  & 37.3  & 3.18e-14  & 0.00848  & 3.9e+05   &  6782  & 1.12e-08  &  0.282 & \\
L2 peak           & 51.8  & 2.71e-14  & 0.00848  & 3.32e+05  &  6260  & 1.04e-08  &  0.261 & \\
                  & 51.8  & 4.41e-14  & 0.00848  & 5.42e+05  &  7992  & 1.32e-08  &  0.333 & \\
1                 & 72.3  & 3.78e-14  & 0.00511  & 4.64e+05  &  9532  & 1.58e-08  &  0.397 & I-front \\
                  & 72.3  & 6.16e-14  & 0.00511  & 7.56e+05  & 12170  & 2.02e-08  &  0.507 & \\
1b                & 49.4  & 2.58e-14  & 0.00511  & 3.17e+05  &  7879  & 1.31e-08  &  0.328 & I-front \\
                  & 49.4  & 4.21e-14  & 0.00511  & 5.17e+05  & 10060  & 1.67e-08  &  0.419 & \\
2                 &  215  & 1.12e-13  & 0.0151   & 1.38e+06  &  9564  & 1.58e-08  &  0.398 & Cometary I-front \\
                  &  215  & 1.83e-13  & 0.0151   & 2.25e+06  & 12210  & 2.02e-08  &  0.508 & \\
2b                &  158  & 8.24e-14  & 0.0105   & 1.01e+06  &  9805  & 1.62e-08  &  0.408 & 2nd I-front \\
                  &  158  & 1.34e-13  & 0.0105   & 1.65e+06  & 12520  & 2.07e-08  &  0.521 & \\
3                 & 55.2  & 2.88e-14  & 0.00785  & 3.54e+05  &  6715  & 1.11e-08  &   0.28 & Tilted I-front \\
                  & 55.2  & 4.7e-14   & 0.00785  & 5.77e+05  &  8574  & 1.42e-08  &  0.357 & \\
4                 &  103  & 5.36e-14  & 0.0075   & 6.58e+05  &  9366  & 1.55e-08  &   0.39 & I-front \\
                  &  103  & 8.73e-14  & 0.0075   & 1.07e+06  & 11960  & 1.98e-08  &  0.498 & \\
5                 &  133  & 6.95e-14  & 0.00617  & 8.54e+05  & 11770  & 1.95e-08  &   0.49 & I-front  \\
                  &  133  & 1.13e-13  & 0.00617  & 1.39e+06  & 15030  & 2.49e-08  &  0.626 & \\
T                 &  104  & 5.43e-14  & 0.00754  & 6.67e+05  &  9407  & 1.56e-08  &  0.392 & Top of $\pi$ \\
                  &  104  & 8.86e-14  & 0.00754  & 1.09e+06  & 12010  & 1.99e-08  &    0.5 & \\
6                 & 40.8  & 2.13e-14  & 0.00785  & 2.62e+05  &  5773  & 9.56e-09  &   0.24 & WR 4083 shell ? \\
                  & 40.8  & 3.47e-14  & 0.00785  & 4.27e+05  &  7371  & 1.22e-08  &  0.307 & \\
7                 & 17.4  & 9.09e-15  & 0.00903  & 1.12e+05  &  3516  & 5.82e-09  &  0.146 & Outer \Hii\ \\
                  & 17.4  & 1.48e-14  & 0.00903  & 1.82e+05  &  4489  & 7.44e-09  &  0.187 & \\
8                 & 13.5  & 7.05e-15  & 0.0208   & 8.66e+04  &  2040  & 3.38e-09  & 0.0849 & Left edge \\
                  & 13.5  & 1.15e-14  & 0.0208   & 1.41e+05  &  2605  & 4.32e-09  &  0.108 & \\
  9 &   14  & 7.31e-15  & 0.0102  & 8.98e+04  &  2966  & 4.91e-09  &  0.124 & \\ 
    &   14  & 1.19e-14  & 0.0102  & 1.46e+05  &  3787  & 6.27e-09  &  0.158 & \\ 
 10 &    8  & 4.18e-15  & 0.0102  & 5.13e+04  &  2242  & 3.71e-09  & 0.0934 & \\ 
    &    8  & 6.81e-15  & 0.0102  & 8.37e+04  &  2863  & 4.74e-09  &  0.119 & \\ 
 11 &    4  & 2.09e-15  & 0.0314  & 2.57e+04  & 903.8  & 1.5e-09  & 0.0376 & \\ 
    &    4  & 3.41e-15  & 0.0314  & 4.18e+04  &  1154  & 1.91e-09  & 0.0481 & \\ 
 12 & 13.5  & 7.05e-15  & 0.0314  & 8.66e+04  &  1660  & 2.75e-09  & 0.0691 & \\ 
    & 13.5  & 1.15e-14  & 0.0314  & 1.41e+05  &  2120  & 3.51e-09  & 0.0883 & \\ 
 13 & 3.01  & 1.57e-15  & 0.0314  & 1.93e+04  &   784  & 1.3e-09  & 0.0326 & \\ 
    & 3.01  & 2.56e-15  & 0.0314  & 3.15e+04  &  1001  & 1.66e-09  & 0.0417 & \\
\hline
	\end{tabular}
\end{table*}

\subsection{Sgr C $HII$ region parameters}

We first discuss the average properties of the Sgr C \Hii\ region,  measured by summing the background-subtracted \Bra\ flux from the portion of Sgr~C in the JWST NIRCam field.  An elliptical region with semi-minor and semi-major radii of 74\arcsec\ and 103\arcsec\ was used for this measurement.  For estimation of the electron density, we use 2-times the semi-minor radius as an estimator of the line-if-sight (LOS) depth  of the \Hii\ region.   Thus, $\rm L_{ii}$=148\arcsec\ or 5.83 pc.  The harmonic mean radius, $\rm R_{ii} \sim $87\arcsec\ (3.45pc),  was used to estimate the mass of plasma and the hydrogen-ionizing Lyman-alpha luminosity of the \Hii\ region.   

We use the continuum-subtracted \Bra\ image to estimate the mean value of the pixel counts in the 148\arcsec\ by 206\arcsec\ diameter elliptical region, finding a mean surface brightness value of $\sim$25.2 MJy~sr$^{-1}$.    Faint \Bra\ emission with a mean value of about $\sim$3 MJy~sr$^{-1}$ is seen throughout the JWST field beyond the suspected, projected boundaries of the Sgr C \Hii\ region.  The background plus foreground contribution to the flux, measured in corners of the \Bra\ image and towards the opaque Sgr C cloud, is $\sim$3 MJy~sr$^{-1}$.  The MeerKAT 1.28 GHz image and the extension of the oval region beyond the JWST field shows that the \Bra\ image misses about 12\% of the projected area of the \Hii\ region. However, the exclusion of this field does not contribute to the mean counts in the measurement oval.  Comparison of the MeerKAT and  \Bra\ images shows that the Sgr~C cloud obscures about 12\% of the area and flux in the measurement oval.  Thus, to get a better estimate of expected mean flux from the \Hii\ region we scale be the background-subtracted mean value (25.2 - 3 = 22.2 MJy~sr$^{-1}$)  by a factor of 1.12, yielding a mean flux density in the oval, 24.9 MJy~sr$^{-1}$.   

We also used a measurement box that avoids the Sgr C molecular cloud and associated extinction, but covers a similar area as the measurement oval,  yielding a background-subtracted mean flux density of 29.6 MJy~sr$^{-1}$.  We use the mean of these two measurement methods to estimate the raw (reddened) flux and use the difference as an estimator of the systematic error.   
Thus, the best estimate of the raw, mean surface brightness density, from the Sgr C \Hii\ region, without correction for foreground extinction,  is 27.2$\pm $2.4 MJy~sr$^{-1}$.  The subtraction artifacts surrounding bright stars, and the contribution of residual emission from fainter stars renders this estimate uncertain by about 10\% , smaller than the 20\% to 30\% error for the small measurement regions used for the filaments and ionization fronts.

For all features within the \Hii\ region, we assume that the foreground K$_s$ (2.15 \mm ) extinction is either  2 or 3  magnitudes \citep{nogueras-lara24}. Thus, this mean surface brightness density is scaled by a factor of 2.65 or 4.32 before being converted into SB units. The emission measure across the entire \Hii~region is  
$\rm EM \approx 1.75 \times 10^5~\rm cm^{-6}~pc$, for A(K$_s$) = 2 magnitudes or 
$\rm EM \approx 2.85 \times 10^5~\rm cm^{-6}~pc$, for A(K$_s$) = 3 
magnitudes.   Thus, the estimated mean electron density is $\rm n_e \approx (EM/L_{ii})^{1/2}$ =
173~\cmq\ to 221~\cmq\ for A(K$_s$) = 2 to 3 magnitudes, lower than the 300~\cmq\ quoted by \citet{simpson18a}. 

For an effective radius of $\rm R_{ii} $ = 3.45 pc, the \Hii\ region plasma mass is about between 992~\Msol\ to 1266 \Msol\ (for A(K$_s$) = 2 to 3 mag.).   The Lyman continuum luminosity required to maintain photo-ionization equilibrium in the \Hii\ region is estimated to be
$\rm Q \approx 0.60~to~0.97\rm \times 10^{50}$~H-ionizing photons per second in the absence of significant dust absorption between the ionizing stars and the outer boundary of the \Hii\ region.  The plasma $\beta$ for the \Hii\ region as a whole is 0.0072 (0.0092 for A(K$_s$) = 3 mag.) assuming a mean field strength of 1 mG.

Table \ref{tab:UCHII} summarizes the properties of the four suspected compact \Hii\ regions \citep[][\citetalias{crowe2024JWST}]{lu19b} that were discussed above and marked in the figures.   

We derive the same parameters from the MeerKAT 1.28 GHz image under the assumption that all the emission originated from thermal bremsstrahlung  (the free-free)  process.   For a thermal plasma, the flux density in the long-wavelength (Rayleigh-Jeans) limit is given by 
$$
S_{\nu} = {{2 k T_{ii} \nu ^2 } \over {c^2}} \tau _{\nu}
$$
where 
$\rm S_{\nu }$ is the total flux density in Jy,
$\rm T_{ii}$ is the plasma temperature of the \Hii\ region,
$\nu$ is the frequency, and $\tau _{\nu}$ is the free-free optical depth 
given by 
$$
\tau _{\nu} \approx 3.28 \times 10^{-7}~ T_4^{-1.35} ~
\nu _{GHz}^{-2.1}  ~EM
$$
where $\rm T_4$ is the plasma temperature in units of $\rm 10^4$~K, 
$\nu _{GHz}$ is the frequency in units of 1 GHz, 
\citep{MezgerHenderson1967,CondonRansom2016}.
Thus, the flux density and emission measure are related by 
$$ \rm
EM = 1.25 \times 10^5 ~T_4^{0.35} ~ \nu _{GHz}^{0.1} ~S_{\nu}
$$

The flux density of the continuum on the MeerKAT 1.28 GHz images in the same oval region used above is $\rm S_{1.28}$=6.68 Jy.   The emission outside the oval region is used to estimate a background surface brightness of about 0.82~Jy in the same area, yielding an estimate for the 1.28 GHz flux from the Sgr~C \Hii\ region of $\rm S_{1.28}$ = 5.86~Jy.    Thus, the emission measure is $\rm EM_{1.28} = 6.3 \times 10^5 ~ (cm^{-6}~pc)$.
Using the same LOS path length as above gives an electron density estimate of 
$\rm n_e (1.28 GHz)$ = 330~\cmq , a plasma mass of about 1915~\Msol\ and a Lyman continuum luminosity $\rm Q \approx 2.13 \times 10^{49}$~s$^{-1}$ .  The resulting electron densities and masses are about a factor of 1.5 to 1.9 times higher than estimated from the extinction-corrected \Bra\ flux.  Given the estimated uncertainties, this is a significant difference.   Either the extinction is underestimated, or there is a contribution of non-thermal emission to the 1.28 GHz flux.  The latter conclusion is consistent with 
the non-thermal spectral index at the MeerKAT frequency of 1.28 GHz.   

The masses of plasma associated with the ionization fronts and filaments are small.  For the densest features, the mass within a spherical region whose diameter is comparable to the projected width of the feature is typically less than $\rm 2 \times 10^{-4}$~\Msol .  At an electron density $\rm n_e = 10^4$~\cmq\ a 1 square arcsecond region would have  mass of order 0.02~\Msol .
A key result from this analysis is that for a typical (large-scale average) magnetic field strength of 1 mG, even the brightest and most compact features seen in the JWST images are magnetically dominated, with a plasma $\beta$ less than one.

\subsection{Spectral indices}\label{sec:spectral_indices}

To investigate the physical origin of these filaments, the spectral index was measured in the Sgr C \Hii~region.  Because of the varying resolution and spatial filtering in the ALMA and MeerKAT data,  we measured the spectral indices of compact structures and filaments between 1.28~GHz and 97 GHz by removing the estimated contributions of the extended or unrelated foreground and background low-frequency emission.   We selected three measurement boxes in the MeerKAT images for this background removal.  `Back1' is located just beyond the outer edge of the \Hii\ region, as seen in the 97 GHz ALMA image, but inside the extended envelope of 1.28 GHz emission surrounding Sgr~C region. `Back2' is a smaller box inside the bright core of the \Hii\ near the $\pi$-shaped filaments, but avoiding any of these filaments.  `Back3' is located even farther from Sgr C in a low emission region in the MeerKAT image, and samples the general background and foreground 1.28~GHz emission in the CMZ and the Galactic disk.  Back1 and Back3 are shown in the top panel of Figure \ref{fig:mediumfield_meerkat}; Back2 is shown in the bottom panel of Figure \ref{fig:bra_zoomin}.    

Table~\ref{tab:spectral_indices} presents the spectral index measurements between the 97 GHz (3 mm) ALMA continuum image and the 1.28 GHz MeerKAT image in the regions indicated in Table \ref{tab:regions}.   The columns labeled SB(97) and SB(1.3) give the 97 GHz and 1.38 GHz mean surface brightness in each region in \mJyas\ without any background subtraction.   The column $\alpha _{raw}$ gives the spectral indices computed from these fluxes.   The presence of 1.28 GHz emission throughout the entire CMZ implies that a better estimate of the spectral index for small features localized in the Sgr~C \Hii\ region requires removal of a background.    We used the three regions described above to estimate this background.  Columns SB1(1.3), SB2(1.3), and SB3(1.3) are surface brightness estimates of compact regions (in \mJyas\ units) after subtracting the mean surface brightness values in regions `Back1', `Back2', and `Back3', respectively.   Columns $\alpha _1$ through $\alpha _3$ give the corresponding spectral index values between 97 GHz and 1.28 GHz for SB1(1.3) through SB3(1.3).

\begin{table*}
	\centering
	\caption{\label{tab:UCHII} Physical parameters of compact \Hii\ regions }
	\label{tab:UCHII}
	\begin{tabular}{lcccccccl} 
		\hline
Region	& counts &    Corrected SB               & L    &  EM                & $\rm n_e$ &   M      &    Q           & Comments \\
  &  (MJy/sr) & ($\rm erg~s^{-1}cm^{-2}as^{-2}$) & (pc) & ($\rm cm^{-6} pc$) & (\cmq )  & (\Msol )  & ($\rm s^{-1}$) & \\
\hline
UC 1  &  69  & 0.6$\times 10^{-13}$  & 0.027  & 0.72$\times 10^6$  &  5180  & 0.0017  & 3.2e45 & Cometary \\
mm1   & 160  & 1.4$\times 10^{-13}$  & 0.021  & 1.67$\times 10^6$  &  8856  & 0.0015  & 4.6e45 &  Barely resolved \\ 
mm2   & 137  & 1.2$\times 10^{-13}$  & 0.045  & 1.43$\times 10^6$  &  5640  & 0.0090  & 1.8e46 & Extended halo \\ 
UC 2  &  54  & 0.5$\times 10^{-13}$  & 0.045  & 0.57$\times 10^6$  &  3541  & 0.0056  & 6.9e45 &  Extended halo\\ 
		\hline
	\end{tabular}
Notes:  The errors on SB and EM are estimated to be about 20\% to 30\%, due to the choice of apertures used to measure the compact \Hii\ region fluxes.  An extinction of K$_s$ = 3 magnitudes is assumed. If the extinction is higher, the values presented here are lower bounds. 
\end{table*}

\begin{table*}
	\centering
	\caption{\label{tab:spectral_indices} Spectral indices of bright features from 98 GHz to 1.28 GHz.  }
	\begin{tabular}{cccccccccc} 
		\hline
Component & SB(97) & SB(1.3)  & SB1(1.3) & SB2(1.3) & SB3(1.3) & $\alpha _{raw}$ & $\alpha _{1}$ & $\alpha _{2}$ & $\alpha _{3}$ \\
    &(\mJyas )&(\mJyas )&(\mJyas )&(\mJyas )&(\mJyas )&      &       &      & \\
        \hline
Leg 1  & 0.182 & 0.781 & 0.531 & 0.233  & 0.755 & -0.336 & -0.247 & -0.057 & -0.328 \\ 
Leg 1b & 0.208 & 0.794 & 0.544 & 0.246  & 0.768 & -0.309 & -0.222 & -0.039 & -0.301 \\ 
Leg 2  & 0.256 & 0.774 & 0.524 & 0.226  & 0.748 & -0.255 & -0.165 & 0.029  & -0.247 \\ 
Top    & 0.458 & 0.981 & 0.731 & 0.433  & 0.955 & -0.176 & -0.108 & 0.013  & -0.169 \\ 
1      & 0.412 & 0.657 & 0.407 & 0.109  & 0.631 & -0.108 & 0.003  & 0.306  & -0.098 \\ 
2      & 0.813 & 0.528 & 0.278 & -0.020 & 0.502 & 0.100  & 0.248  & -      & 0.111  \\ 
3      & 0.459 & 0.872 & 0.622 & 0.324  & 0.846 & -0.148 & -0.070 & 0.080  & -0.141 \\ 
4      & 0.372 & 0.693 & 0.443 & 0.145  & 0.667 & -0.143 & -0.040 & 0.218  & -0.135 \\ 
5      & 0.776 & 1.258 & 1.008 & 0.710  & 1.232 & -0.112 & -0.061 & 0.020  & -0.107 \\ 
Back 1 &    -  & 0.250 & 0.000 & -0.298 & 0.224 & -      & -      & -      &  -     \\ 
Back 2 &    -  & 0.548 & 0.298 & 0.000  & 0.522 & -      & -      & -      &  -     \\ 
Back 3 &    -  & 0.026 & -0.224& -0.522 & 0.000 & -      & -      & -      &  -     \\ 
		\hline
	\end{tabular}
\end{table*}

Table~\ref{tab:spectral_indices} shows that the radio spectral indices of brightest filaments in the JWST \Bra\ image in the Sgr~C \Hii\ region are dominated by free-free emission.  Although no background subtraction, or using backgrounds in regions `Back1' and `Back3', located outside the \Hii\ region, yield negative spectral indices indicative of strong non-thermal emission, use of the `Back2' region (which is inside the \Hii~region; Figure \ref{fig:bra_zoomin}) for background subtraction to estimate the 1.3 GHz radio flux gives an index close to -0.1, which is expected in the case of optically thin, free-free (i.e., purely thermal) emission \citep{Draine11}.  

The filaments labeled 1 through 5 form a fragmented, concentric arc of emission wrapping around the Sgr~C cloud with a radius of $\sim$45\arcsec\ ($\sim$1.85 pc). Given that, with background subtraction, the spectral indices between 97 GHz and 1.28 GHz are close to flat---slightly negative using the flux in `Back 1' and slightly positive using the mean flux in `Back 2'---it is likely that the numbered features trace ionization fronts dominated by thermal free-free emission with some contribution of emission from warm dust.   Although the radio continuum emission between 97 and 1.28 GHz is dominated by the free-free process, the negative spectral index in the 1 to 2 GHz region (i.e. in the MeerKAT spectral index map; see Figure \ref{fig:mediumfield_meerkat}) indicates the presence of some non-thermal component, likely synchrotron radiation.   We therefore hypothesize that the unusual filamentary morphology of Sgr~C results from confinement of the \Hii\ region plasma by strong magnetic fields. 

\subsection{A simple model of filamentary, magnetically dominated \Hii~regions} \label{sec:filament_model}

The filamentary morphology of the MeerKAT, ALMA, and NIRCam data can be explained by confinement of thermal plasma  by strong magnetic fields.  We present a simple model for the evolution of \Hii\ plasma in highly structured, strong magnetic fields, which can explain the unique appearance of the Sgr C \Hii~region.  This picture may explain filamentary  structure of the Central Molecular Zone \Hii\ regions in general, and therefore would be applicable to photo-ionized plasmas in all galactic nuclei.   


Magnetic fields play pivotal roles in star formation \citep{Hennebelle2019,Pattle2019,Pattle2023}.   However, when massive stars form in the Galactic disk, the thermal pressure of photo-ionized plasma tends to dominate the evolution of their \Hii\ regions. A number of studies have explored the roles of magnetic fields in the formation and evolution of \Hii\ regions. \citet{krumholz07} integrated a magneto-hydrodynamic (MHD) code into the existing hydrodynamic framework for modeling \Hii\ region evolution.  
\citet{Soler2017} present models of cloud 
collapse and evolution in the presence of magnetic fields.
\citet{Ibanez_Meja2022} explore the relationship between field orientations, density gradients and flows.  

\citet{arthur11} and \citet{zamoraaviles19} found that magnetic fields inhibit \Hii\ region expansion perpendicular to the field lines but allow expansion parallel to the field lines.   
As  \Hii\ regions grow,  this leads to anisotropic development. \citet{mackey11} present MHD simulations of pillar formation at \Hii\ region edges under uniform B-fields up to $160 ~\mathrm{\mu G}$. B-fields perpendicular to the pillar axes tend to create ribbons of plasma expanding from the tip of the pillar at right angles.   Parallel fields tend to produce ribbons of plasma expanding back into the \Hii\ region parallel to the pillar axis.  In highly magnetized environments such as the CMZ, these trends
will tend to be amplified.
These authors suggest that in highly magnetized \Hii\ regions, such filamentary structures would be readily observable in plasma tracers (e.g., recombination lines such as \Bra~at $4.05~ \mu m$).

Most literature on \Hii\ region evolution in a magnetized medium considers relatively weak ($\leq 50 ~ \mathrm{\mu G}$) fields typical of the Galactic disk star forming regions \citep[see, e.g.,][]{vaneck11}.  Furthermore, most works assume uniform initial field configurations \citep{krumholz07,arthur11,mackey11,zamoraaviles19}. The magnetic fields in the CMZ are orders of magnitude stronger than in the Galactic disk near the Sun, with CMZ field strengths ranging from 100 $\mu $G to over 1~mG \citep{ferriere09}.  The hundreds of  NTFs indicate that the fields are not uniform.  They tend to be confined to ropes or sheets (see, e.g., Figs.\,\ref{fig:widefield_meerkat} and \ref{jwst_alma_meerkat}). Nonetheless, the formation of filaments in strongly magnetized \Hii\ regions is suggested in previous literature; as mentioned, \citet{mackey11} considered \Hii\ region evolution in regions with magnetic fields of $\sim 200 \mathrm{ ~\mu G}$, finding the formation of filamentary striations in the plasma. It is likely that in even stronger and anisotropic magnetic fields, such as those in the CMZ, the formation of filamentary structure in \Hii\ regions would be even more pronounced. The morphology of the Sgr C \Hii\ region provides direct observational evidence of this.

We propose that the CMZ \Hii\ regions evolve in a highly magnetized medium, and that this environment produces the filaments seen in Sgr C. The plasma $\beta$ (thermal pressure dominates over magnetic pressure; Eq. \ref{eq:plasma_beta}) is large in Galactic plane \Hii\ regions \citep[based on typical Galactic disk field strengths;][]{crutcher12}. On the other hand, the plasma $\beta$ in the CMZ may be very low (magnetic pressure dominates over thermal pressure), as shown below.

In a high $\beta$ regime, the expansion of \Hii\ regions is driven by thermal pressure (Eq. \ref{eq:thermal_pressure}). In a low $\beta$ environment, the magnetic pressure (Eq. \ref{eq:magnetic_pressure}) will dominate. Using an electron density of $\rm n_{\sc ii} = 300$~\cmq\ \citep{simpson18a} and a plasma temperature of $T_{\sc ii} = 6,000$~K gives a thermal pressure $\rm P_{\sc ii} \approx 5 \times 10^{-10}$
dynes~\cms\ for the Sgr~C \Hii\ region.  At this density, the plasma $\beta$ will be less than 1 for a field stronger than 79 $\mu$G.   For a 1 mG field, typical of dense CMZ molecular clouds \citep{ferriere09,pillai15}, $\rm P_B \approx 4 \times 10^{-8}$ and $\rm \beta \sim 1.2 \times 10^{-2}$.   In the high $\beta$ (i.e. thermally dominated) limit,  \Hii\ region evolution will proceed according to the standard Oort-Spitzer description \citep{Draine11}, modified by the density structure of the environment and the actual evolution of the stellar Lyman continuum luminosity.  As the \Hii\ region density and pressure drops (e.g. due to outward expansion) in a highly-magnetized medium,  $\beta$ declines and eventually enters the low-$\beta$ regime, after which the structure of the magnetic field will direct further nebular evolution.    In an ideal,  uniform density, uniform field environment, the initial, high-$\beta$ expansion would displace the field, trapping it in a D-type (dense shell) ionization front. As $\beta$ drops below one, the \Hii\ region would preferentially continue to expand along the field lines without significant growth at right angles to the field.

If a magnetized rope or sheet is loaded with neutral gas below a critical density,  $\rm n_{crit}$,  Lyman continuum radiation will drive an R-type (rapid) ionization front through the cloud with a speed faster than twice the sound speed in the plasma.  The abrupt increase in temperature will raise the thermal pressure, causing the field and trapped plasma to expand.   This critical density of neutral gas (in cm$^{-3}$) is given by 
\begin{equation}\label{eq:crit_density}
\rm n_{crit} \approx (3 Q / (4 \pi \alpha_B))^{1/2} R^{-1/2} D^{-1}
\end{equation}
where $\rm Q$ is the Lyman continuum luminosity of a star (or group of stars)
located at a distance $\rm D$ from a magnetized filament with radius $\rm R$, or
a sheet with half-thickness $\rm R$,  and 
$\rm \alpha_B \approx 2.6 \times 10^{-13} T_4^{-0.83}$~$\rm cm^{-3} s^{-1}$ 
is the case-B recombination coefficient for hydrogen with the plasma temperature
measured in units of $\rm T_4 = 10^4$~K \citep{Draine11}.  
 
If the density in the magnetized region is much greater than $\rm n_{crit}$, but the surrounding regions have a lower density, the ionization front will race around the dense region, but stall when it hits the dense gas.   The increased pressure behind the ionization front will drive a D-type (dense) shock front that jumps out ahead of the ionization front and compresses the medium.  As the surface layers of a magnetized filament are ionized, magnetic fields and plasma are peeled off.  But, the ram pressure of this magnetized photo-ablation flow will tend to compress the colder neutral gas in the rope or sheet, as well as the magnetic field.   

This picture differs from the standard description of propagating D-type ionization fronts in several ways.  First, soft, non-ionizing UV will pre-heat the neutral medium, weakening the shock that would, in the isothermal limit, compress gas prior to becoming ionized. Second, the magnetic field will also resist compression. The magnetic field pressure may eliminate the D-type shock discontinuity altogether by converting it into a so-called C-type (continuous) structure, in which fluid variables change gradually rather than abruptly.   Third, ionized plasma will preferentially flow along the field lines, forming filaments aligned with the field, rather than orthogonal to the ionization fronts as in the standard, non-magnetized picture of photo-ablation.

\section{Discussion}
\label{sec:discussion}

Among visual and near-infrared emission nebulae, filamentary structure tends to be associated with shock fronts, ionization fronts, and photon-dominated regions (PDRs).  Filaments generated by shocks are commonly seen in supernova remnants (SNRs), some stellar wind bubbles, planetary nebulae,  various types of outflows from stars, and superbubbles.  Examples include the relatively nearby Crab Nebula, Cas A, Veil Nebula, and Simeis~147 supernova remnants, the NGC~2359 (`Thor's Helmet')  stellar wind bubble, protostellar bow shocks such as those in HH~46/47, and the Orion-Eridanus superbubble.   The Mittelman, di Cicco, and Walker (MDW) \Ha\ survey provides one of the deepest, wide-field surveys of the local ISM in a hydrogen recombination line (see https://www.mdwskysurvey.org).    These ultra-deep \Ha\ images of the local ISM (within about 1 kpc of the Sun) show extensive, faint filaments, and have led to the discovery of new, but faint, supernovae remnants \citep{Fesen2024}.    Filaments such as those tracing shocks moving at right angles to our line-of-sight are limb-brightened because the line-of-sight is nearly tangent to the shock front.   When such filaments move into a uniform medium they can be relatively straight.   But, the filaments in the post-shock environment tend to present a chaotic jumble of curved structure.  The Crab, Veil and Simeis~147 SNR are good examples.   Ionization fronts and PDRs tend to have a softer appearance and are often usually associated with dramatic increases in extinction as one moves from the ionized \Hii\ region or soft-UV dominated bubble into the surrounding neutral medium.  Ionization fronts and PDRs tend to exhibit highly clumpy structure.

It is possible that some filaments seen in the Sgr C JWST field, especially those that lie beyond the projected outer boundary of the Sgr C \Hii\ region, are shock waves similar to those seen in supernova remnants and stellar wind bubbles near the Sun.  Alternatively,  the fainter filaments outside the projected edge of the \Hii\ region may trace magnetically confined plasma ionized by OB stars outside Sgr C.  These filaments may be located anywhere along the line-of-sight through the CMZ.   Although some filaments in the projected interior of the Sgr C \Hii\ region may also be foreground or background objects, the majority are likely ionized by the Sgr C OB and WR stars.  Evidence for this claim is based on the high surface brightness of these features compared to those outside the \Hii\ region.  However, since there are two WR stars in Sgr C, it is possible that this \Hii\ region has been processed by stellar winds bubbles and possibly by supernovae.    The linear morphology of these filaments makes it unlikely that most owe their origins to such shocks.   Hence, we propose that magnetized sheets or ropes are a more likely cause of this morphology.

The approximately flat Galactic rotation curve within a few-hundred pc of the CMZ \citep{sormani20,sormani22}  will tend to drive a shear dynamo, resulting in the dominance of the toroidal component of the magnetic field in the dense gas, traced by polarized sub-mm emission of the dust \citep{Mangilli2019}.  However,  the majority of NTFs show that there is a poloidal component to the field \citep{morris07}.   Bubble expansion driven by \Hii\ regions, stellar winds, supernovae, and superbubbles may lift field-lines and plasma away from the dense gas near the mid-plane of the Galactic disk, where most of the stellar feedback sources were born. Since, in the mid-plane, the bubble walls tend to be orthogonal to the Galactic plane, fields swept-up into the bubble walls (appearing as e.g. NTFs) will also tend to be orthogonal and trace a poloidal field component \citep{sofue23}.    
The `Radio Arc' near l$\sim 0.18\arcdeg$, which is at the high-longitude edge of a compact superbubble, and spatially associated with the massive Arches and Quintuplet clusters (as well as `The Sickle', a curved \Hii~region), is a prominent example of this phenomenon \citep{heywood22}. Hubble Space Telescope Paschen-$\alpha$ data in the vicinity of this region, presented in \citet{dong11}, also show filaments in CMZ \Hii\ regions located between Sgr A* and the Arches and Quintuplet clusters.

The large random motions of CMZ clouds superimposed on their ordered circulation about the Galactic center also tend to tangle magnetic fields.  Instabilities in magnetized shells and sheets may further fragment the fields into magnetic filaments and ropes.   As such, magnetic structures become loaded with relativistic electrons from discrete sources, such as pulsars, or by in-situ particle acceleration by shocks. The resulting synchrotron emission creates NTFs.  The enhanced cosmic ray fluxes in the CMZ \citep[$\zeta \gtrsim 10^{-15}\:{\rm s}^{-1}$;][]{carlson16} guarantee that neutral atomic and molecular gas will be strongly coupled to the magnetic fields.

The Sgr C molecular cloud appears in silhouette against the surrounding \Hii\ region (Figure \ref{jwst_alma_meerkat}).   The absence of obvious ionization fronts at the edges of the cloud, or projected onto its interior, suggests that the clouds lies in front of the ionized region. Additionally, the observed high-dipole moment molecules such as CS, HNCO, etc.  indicate that the mean density of the cloud is $\rm n(H_2) > 10^5$~\cmq\:\citep{nguyenqreiu91}, 
especially towards its cometary head, where the most luminous IR sources and massive protostars are located. 
However, some small amount of \Bra\ emission does leak through the upper part of the cloud
which may trace photo-ionization of the backside.  

As stated in the Introduction, \citet{lu24} used millimeter-wavelength dust polarization measurements to trace the magnetic field orientation in the Sgr~C molecular cloud, finding that the field wraps around the cloud, with the predominant orientation parallel to the cloud surface.   Most \Bra\ filaments near the projected edge of the cometary cloud do not follow this orientation.   Assuming that the cloud surface is being ionized and photo-ablated by Lyman continuum radiation, the thermal pressure at the ionization front may be locally higher than the magnetic pressure.  In this high-$\beta$ environment, the plasma flow will tend to be along the local density and pressure gradient (e.g.  away from the ionization front) roughly at right angles to the front.  The plasma may drag along the magnetic field, stretching it along the flow direction.   This phenomenon may explain some of the filaments that appear point away from the cloud (see, e.g., Figure \ref{fig:bra_zoomin}).  

Figure~\ref{fig:bra_zoomin} shows that several bright, parsec-scale filaments point back towards the cometary head of the  Sgr~C cloud, which contains a cluster of luminous sub-mm cores and forming massive protostars \citep[][\citetalias{crowe2024JWST}]{kendrew13,lu20,lu21,lu22}.   Massive protostars tend to drive the most powerful outflows in forming clusters.  It is possible that some of these filaments trace the walls of fossil outflow lobes that emerged from the Sgr C molecular cloud.  Alternatively, it is possible that magnetic ropes threaded the core of the protocluster and the surrounding medium prior to ionization.    
If the lower density medium surrounding the dense cloud was permeated by filaments or sheets of magnetic field prior to being ionized by massive stars, the plasma would flow along the field lines if  the plasma $\beta$ remains well below 1.  
Although the parsec-scale filaments that run nearly horizontally in Figure \ref{fig:bra_zoomin} do exhibit `wiggles' over their full extent, 
the shorter filaments constituting the $\pi$-shaped structure in the \Bra\ image (Figure \ref{fig:Pi_filament}) are straight, with length-to-width ratios ranging from 10 to over 100.   The lack of bends or entanglement may imply that the magnetic pressure in the filaments must be higher than the ram pressures of various flows that would cause deflections.
The tension in the field lines must exceed the amplitudes of these forces.  The large length-to-width ratio suggests that the plasma $\beta$ is below about 0.01.   However, the parameters derived for the brightest and presumably densest filaments in Table 2 have plasma $\beta$
values between 0.3 and 0.6.  These values were based on the assumption that the projected width of a filament is equal to its line-of-sight depth.  The large length-to-width ratios may  indicate that 
these filaments are mostly sheets seen edge-on.  In this case, the assumption that their widths are comparable to their line-of-sight
depths overestimates their electron densities and thermal pressures.

Non-thermal emission has been seen in other CMZ \Hii\ regions.  JVLA observations of the Sgr B2 complex show that the envelopes surrounding the various cluster-forming cores in Sgr B2 exhibit non-thermal spectral indices, especially in the extended, V-shaped region Sgr B2 South, located south of the main Sgr B2 clumps \citep{Meng2019}.    
\citet{Padovani2019} present a model of non-thermal emission from cosmic rays accelerated in \Hii\ regions.

\subsection{Ionizing sources in Sgr C: Wolf-Rayet stars and wind bubbles}\label{sec:wr_stars}

There are two spectroscopically-confirmed Wolf-Rayet (WR) stars in Sgr C, 2MASS J17443734-2927557 and J17444083-2926550 (which we shorten to WCL 3734 and WCL 4083, respectively) \citep{clark21}. They are of spectral type "WCLd" \citep[``L'' stands for ``Late'' on the basis of emission line strengths, and ``d'' stands for ``dust'' evident in the spectrum;][]{crowther07,liermann12}. In general, WCL stars have ionizing photon emission rates of $\rm Q \approx 10^{49}$~H-ionizing photons per second, meaning that these two stars cannot fully account for the estimated ionizing photon emission rate in the region alone ($\rm Q \approx 6~to~10 \times 10^{49}$~H-ionizing photons per second). However, while there are almost certainly additional massive, ionizing stars in the region \citep{nogueras-lara24}, their  locations are unknown.

WCL 3734 and WCL 4083 have been confirmed to lie within the CMZ by \citet{oka19} on the basis of detection of strong absorption of H$_3^+$ in their stellar spectra, both from foreground spiral arms and from warm CMZ gas. WCL 3734 is located in the center of the Sgr C \Hii\ region. It is roughly centered between a pair of parallel filaments separated by about 3.3\arcsec\ ( 0.13 pc) that can be traced for at least 1.5 pc in length (Fig. \ref{fig:Pi_filament}).   We propose that this structure traces a cylinder of magnetized plasma that has displaced by the pressure of a stellar wind from the WR star.  

The WR star J17444083-2926550 (WCL 4083) is located northeast of the Sgr C \Hii\ region core.  WCL 4083 is sandwiched between a group of parsec-scale filaments extending from the surface of the Sgr~C GMC.   These filaments reconnect about one parsec to the upper left of WCL 4083 (towards due North in equatorial coordinates; see Figure \ref{fig:bra_zoomin}). They may trace the walls of a parsec-scale cavity created by the winds of this WR star, similar to WCL 3734.    

Wolf-Rayet stellar winds have typical mass loss rates ranging from $\rm 10^{-7}$ to over $\rm 10^{-4}$~\Mdot\ with speeds ranging from 1,000 to over 4,000 \kms~ \citep{crowther07}.  The interface between the free-flowing WR star wind and the surrounding \Hii\ region contains two shocks.  The forward shock, driven by the ram pressure of the wind, sweeps-up \Hii\ region plasma into a dense shell.  A reverse shock propagates back into the free-flowing wind, which thermalizes the wind kinetic energy and creates an X-ray hot ($\rm 10^7$ to $\rm 10^8$ K) bubble of shocked stellar wind \citep{Castor1975,Weaver1977}.    The pressure between the forward and reverse shocks is determined by the ram pressure of the forward shock.    These fast shocks may be ideal environments for the acceleration of particles to relativistic speeds.  While in an un-magnetized, uniform density medium, stellar winds bubbles are expected to be roughly spherical, in a low-$\beta$ plasma with a strong, organized magnetic field, and with a magnetic pressure that exceeds the ram-pressure of the wind, the field will channel the wind bubble into a cylindrical shape. As an example, consider a wind with a speed of 1,500 \kms\ and a mass-loss rate of $\rm 10^{-6}$~\Mdot\ expected of a WCL star \citep{crowther07}.  At a radius of 0.13 pc (the radius of the cavity surrounding WCL 3734), the wind ram pressure would be about $\rm 5 \times 10^{-9}$~dynes~$\rm cm^{-2}$.    For any field greater than 0.3 mG, the field would dominate the evolution of the wind bubble, funneling it into the cylindrical morphology seen in Figure \ref{fig:Pi_filament}.   

\section{Conclusions}\label{sec:conclusions}

\begin{table*}
	\centering
	\caption{Summary of Sgr C \Hii\ Properties}
	\label{tab:Hii}
	\begin{tabular}{lcccccccl} 
		\hline
Method 	                 &  Raw SB &  Corrected SB & L &  EM  & $\rm n_e $ &   M(\Msol) & $   Q   $ & Comments \\
                    &  (MJy/sr) &    (\SB )   & (pc) & ($\rm cm^{-6} pc$) & (\cmq )  & (\Msol )  & ($\rm s^{-1}$) & \\
\hline
\Bra\ A(K$_s$) = 2 mag.  &  29.27   & 1.53$\times 10^{-14}$ & 5.83 & 1.88$\times 10^{5}$ & 179 & 1028  & 0.64$\times 10^{50}$ & \\
\Bra\ A(K$_s$) = 3 mag.  &  29.27   & 2.49$\times 10^{-14}$ & 5.83 & 3.06$\times 10^{5}$ & 229 & 1313  & 1.00$\times 10^{50}$ & \\
1.28 GHz &  5.86 Jy & 42 $\mu$Jy~as$^{-2}$ & 5.83 & 6.28$\times 10^{5}$ & 328 & 1915  & 2.13$\times 10^{50}$ & Assume f-f \\
		\hline
	\end{tabular}
\end{table*}

We use JWST \Bra\ and nearby continuum images of the Sgr C \Hii\  region to measure its physical parameters.   These are summarized in Table \ref{tab:Hii}.  The intrinsic \Bra\ luminosity and the derived emission measure depend sensitively on the assumed extinction and the treatment of foreground and background emission estimated from portions of the JWST image away from the \Hii\ region.  The measurement of these parameters are limited by the high density of stars and their impact on continuum subtraction.  The background subtraction and stellar confusion result in a roughly 10\% error in the mean surface brightness and emission measure of the entire \Hii\ region (less than the estimated 20\% to 20\% error in background-subtracted flux measurements for compact features).  If the background-subtracted 1.28 GHz MeerKAT flux from the Sgr C \Hii\ region were entirely due to free-free emission,  the derived emission measure,  mean electron density, and mass would be larger by a factor of 1.5 to 1.9.
This discrepancy is an indirect indication that 
a large portion of the MeerKAT emission must be produced by the non-thermal,  synchrotron process.  This conclusion is consistent with the in-band spectral index around 1.28 GHz and provides indirect evidence for a strong magnetic field.    

The CMZ is pervaded by filamentary structure across all length scales, from sub-parsec scales to $\geq\rm 100 pc$.  NIRCam observations of Sgr C has shown that the filamentation in CMZ \Hii\ regions is not only evident in non-thermal emission, but is also strongly coupled to free-free emission from thermal plasma. This provides evidence that the evolution of \Hii\ regions in highly magnetized, anisotropic fields is distinct from the traditional model in the Galactic disk. This has ramifications not only for our understanding of star formation and the flows of mass and energy across the CMZ of our Galaxy, but in all galactic nuclei and any other analogous environments.

The filamentation observed in Sgr C by JWST, and suggested to be ubiquitous across the CMZ by MeerKAT, reveals a pressing need to incorporate the impact of strong magnetic fields into the current understanding of \Hii\ region evolution. It is expected that high-density cores giving birth to massive OB stars may encompass a transition from the high to low $\beta$ regime.  The high accretion rates associated with massive star birth may result in bloated stellar photospheres which are too cool to emit Lyman continuum radiation. As the accretion rate onto the massive star decreases, its photosphere can shrink, heat, and start to emit Lyman continuum.  The environment that becomes ionized is likely to have been pre-processed by protostellar jets and outflows.  Such flows may have dragged magnetic fields, elongating them along the flow direction.   

The plasma created by the ionization of dense gas at the surfaces of in circumstellar disks and the surrounding molecular core and clump may be sufficiently dense so that the total pressure is dominated by thermal pressure and turbulence.  Disks, cores, and clumps in massive star forming regions tend to be warm with T$\sim$100 to 1,000 K. Ionization raises the temperature and thermal pressure by one to two orders of magnitude.  If the thermal and magnetic pressures were in rough equipartition during the warm phase prior to ionization, the plasma $\beta$ of the freshly ionized medium will initially be large.  As the plasma expands due to the suddenly increased pressure and flows into previously generated outflow channels or in funneled by magnetic fields, there may be a transition to the low-$\beta$ (i.e., magnetically dominated) regime as the plasma density decreases. 

There appear to be at least two distinct populations of filaments in the CMZ: the long-studied large-scale ($\sim \rm tens~of~parsecs$), entirely non-thermal (i.e. invisible in recombination emission) filaments, and the parsec- to sub-parsec scale filaments that have both a non-thermal (synchrotron) and thermal (free-free) component that are strongly coupled. We assert that the latter of these two populations of filaments are prevalent in all CMZ \Hii~regions, due to the presence of strong, anisotropic magnetic fields and ionizing radiation. Further observations of CMZ \Hii~regions (e.g., Sgr B1), especially in hydrogen recombination lines such as \Bra~and Paschen $\alpha$, as well as accurate modeling of \Hii~region evolution in CMZ-like environments, will be necessary to probe more of this population of filaments and to affirm the trend of \Hii~region striation observed in Sgr C.


\begin{acknowledgements}
\textit{Acknowledgments:} This work is based on observations made with the NASA/ESA/CSA James Webb Space Telescope. The JWST data presented in this article were obtained from the Mikulski Archive for Space Telescopes (MAST) at the Space Telescope Science Institute. The specific observations analyzed can be accessed via \dataset[DOI: 10.17909/y4e2-k269]{https://doi.org/DOI}.
These observations are associated with program~4147. Support for program~4147 was provided by NASA through a grant from the Space Telescope Science Institute, which is operated by the Association of Universities for Research in Astronomy, Inc., under NASA contract NAS 5-03127. 
This paper makes use of the following ALMA data: ADS/JAO.ALMA~2021.1.00172.L. ALMA is a partnership of ESO (representing its member states), NSF (USA) and NINS (Japan), together with NRC (Canada), NSTC and ASIAA (Taiwan), and KASI (Republic of Korea), in cooperation with the Republic of Chile. The Joint ALMA Observatory is operated by ESO, AUI/NRAO and NAOJ. The National Radio Astronomy Observatory is a facility of the National Science Foundation operated under cooperative agreement by Associated Universities, Inc.

J.B.  acknowledges support by National Science Foundation through grant No. AST-1910393 and AST-2206513. S.T.C. acknowledges support from the award JWST-GO-04147.003-A. R.F. acknowledges support from the grants Juan de la Cierva FJC2021-046802-I, PID2020-114461GB-I00, PID2023-146295NB-I00, and CEX2021-001131-S funded by MCIN/AEI/ 10.13039/501100011033 and by ``European Union NextGenerationEU/PRTR''. A.G. acknowledges support from the NSF under grants AAG 2008101, 2206511, and CAREER 2142300. R.S. acknowledges financial support from the Severo Ochoa grant CEX2021-001131-S funded by MCIN/AEI/ 10.13039/501100011033. MGC, from grant EUR2022-134031 funded by MCIN/AEI/10.13039/501100011033 and by the European Union NextGenerationEU/PRTR,  and from grant PID2022-136640NB-C21 funded by MCIN/AEI 10.13039/501100011033 and by the European Union. Z.-Y.L. is supported in part by NSF AST- 2307199 and NASA 80NSSC20K0533.

\end{acknowledgements}




\bibliography{masterbib.bib}

\newpage

\begin{figure*}
\centering
\includegraphics[width=1.0\textwidth]{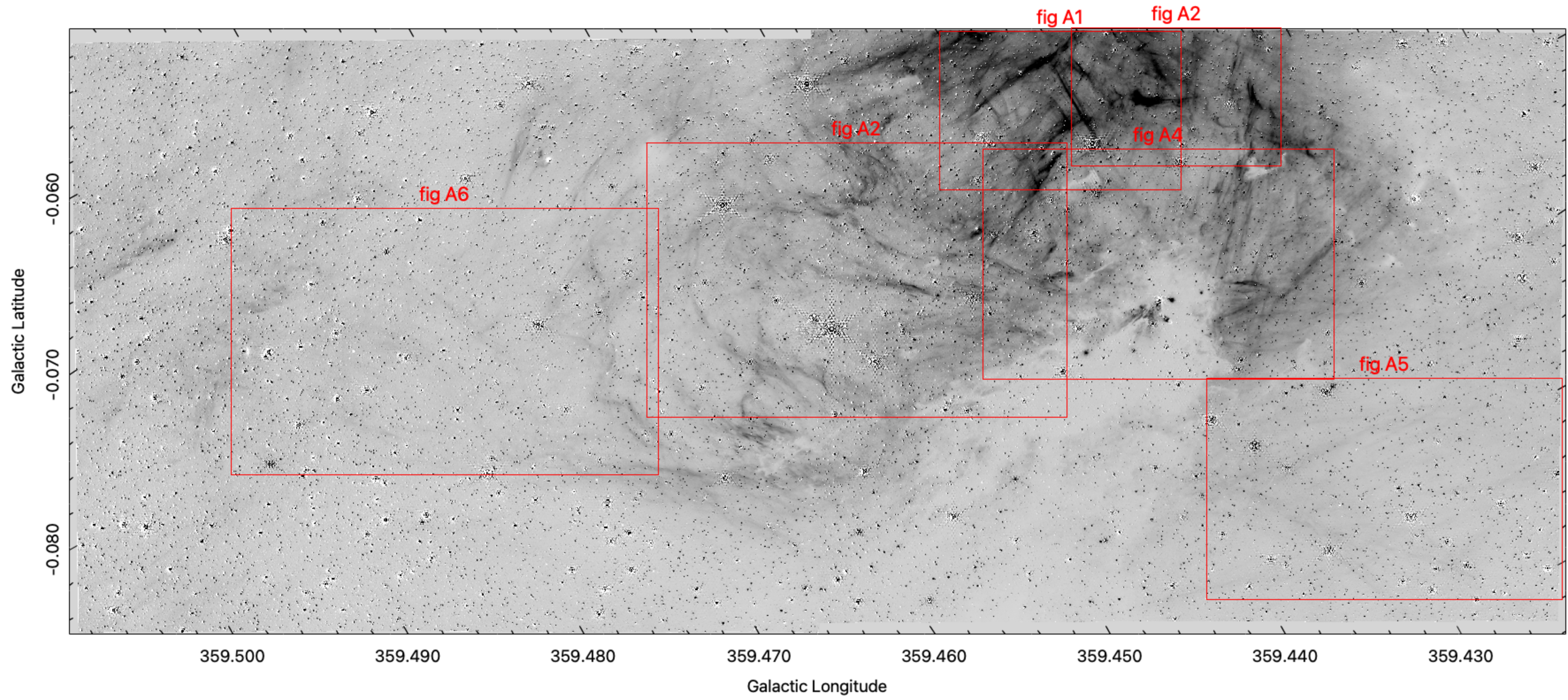}
\caption{\label{figA0} A finder chart showing the locations of the next six
figures, labeled from A1 to A6.}
\end{figure*}
\appendix 
\restartappendixnumbering
\section{Gallery of magnified views of the Sgr C \Hii\:region} \label{section:appendix_magnified_images}
We present close-up views of selected regions in the Sgr C \Hii\ region.
Figure \ref{figA0} shows the entire JWST field on which the locations
of the figures shown in this Appendix are shown.   Note that in these figures
the tick-marks are tilted because the images are shown in the JWST pixel coordinate
space, which, as shown in Figure 1, is tilted with respect to the Galactic 
coordinate system.

\begin{figure*}
\centering
\includegraphics[width=1.0\textwidth]{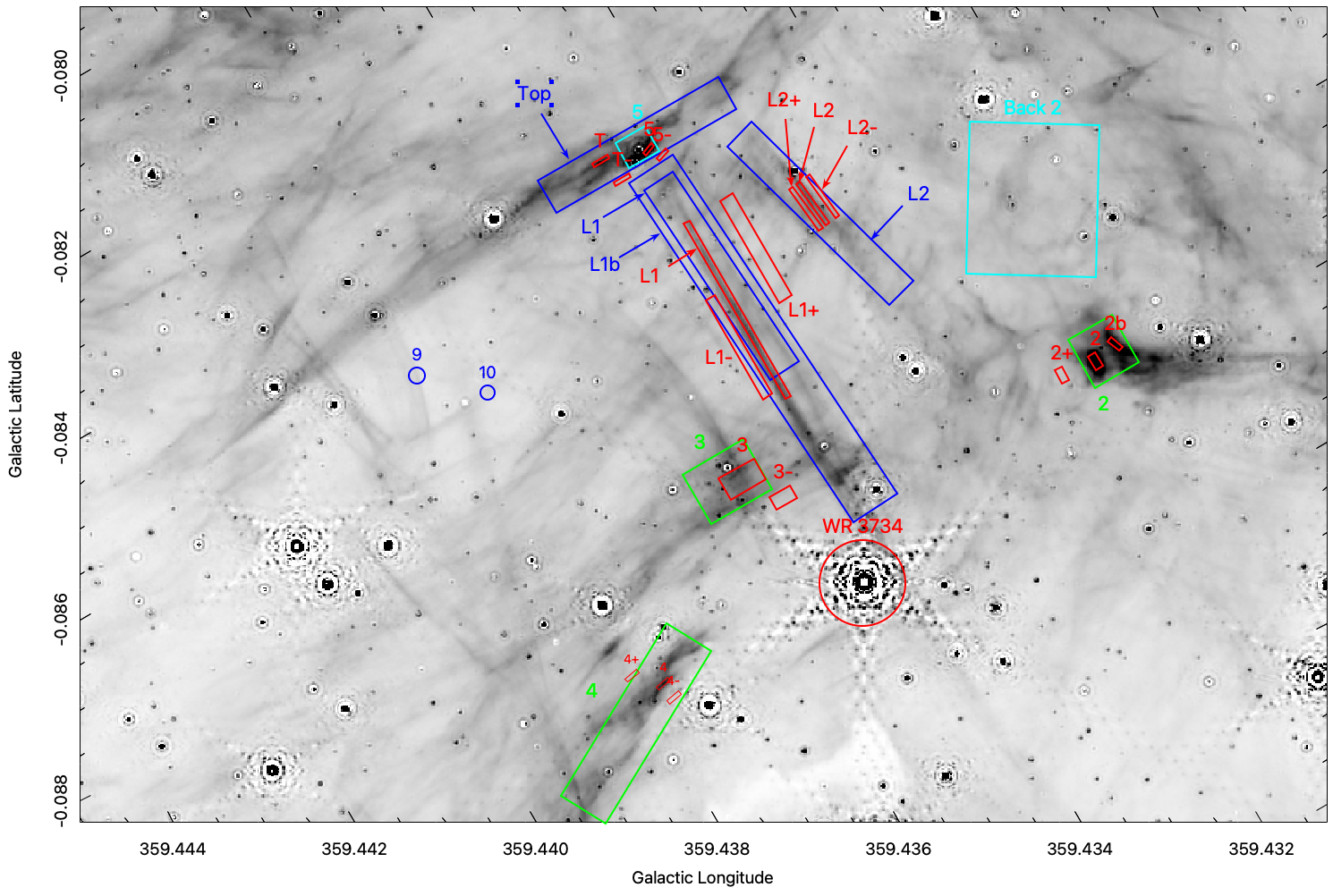}
\caption{\label{figA1} A closeup view of the Br-$\alpha$
emission from the $\pi$-shaped filaments, the cylindrical cavity centered on the Wolf-Rayet star WCL 3734, and the inverted V-shaped filaments to the left of the $\pi$-shaped filaments.   Red boxes show measurement regions used for determining \Bra\ fluxes.}
\end{figure*}

\begin{figure*}
\centering
\includegraphics[width=1.0\textwidth]{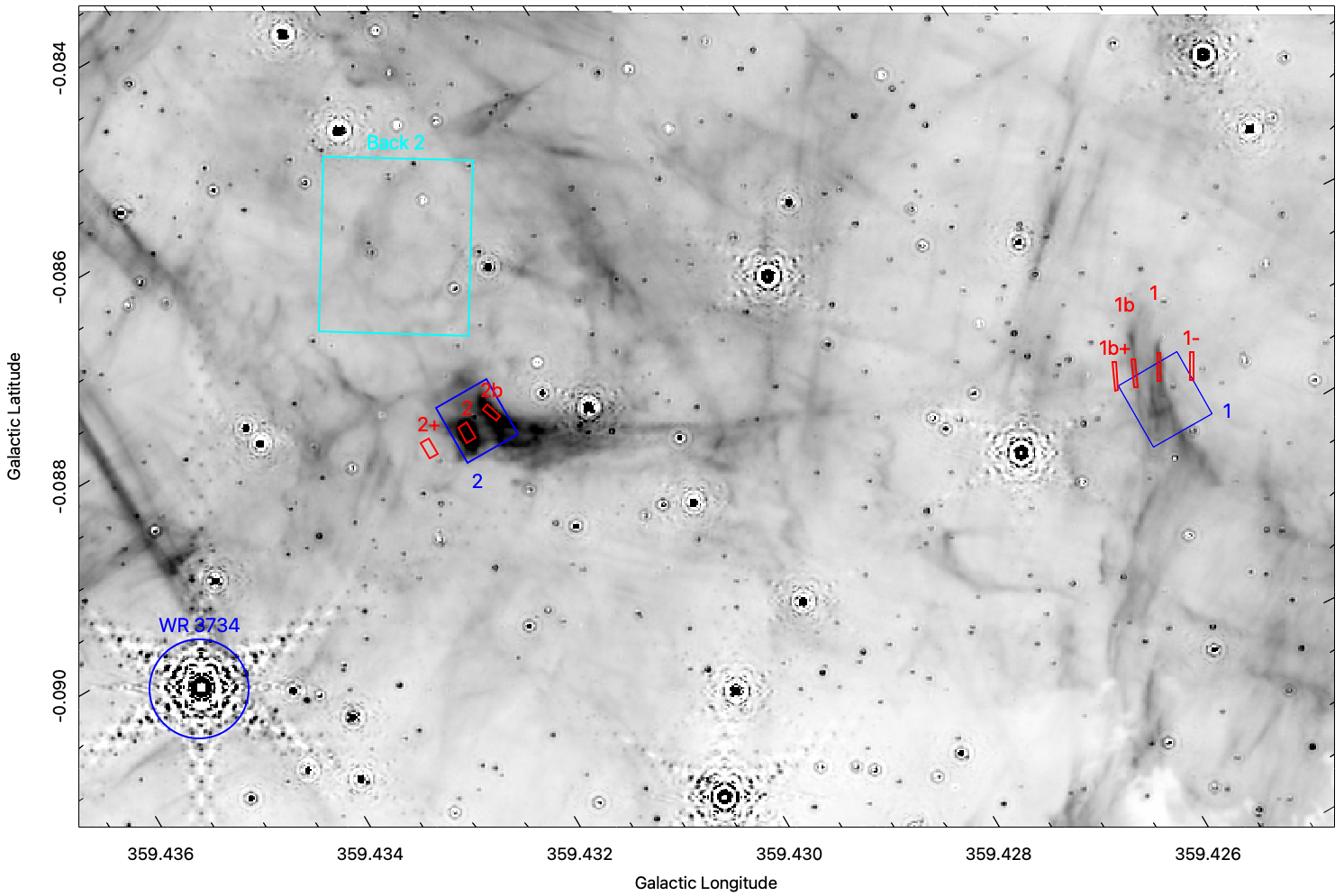}
\caption{\label{figA2} A closeup view of the Br-$\alpha$
emission to the right of the $\pi$-shaped filaments. Red boxes show measurement regions used for determining \Bra\ fluxes.}
\end{figure*}

Figure \ref{figA1} and \ref{figA2} show close-up views of the brightest
portion of the Sgr C \Hii\ region.   In Figure \ref{figA1}, note the chain of
V-shaped structures located between Galactic longitudes 359.440 and 359.444.
Close inspection shows these filaments to be double, with two strands separated by
a few arc-seconds.    These Figures show the regions used to measure \Bra\ fluxes
(numbered boxes) along with regions used for measuring local backgrounds (numbered boxes
with a + and/or - sign).  Note the crosshatching pattern of filaments in Figure 
\ref{figA2}.

\begin{figure*}
\centering
\includegraphics[width=1.0\textwidth]{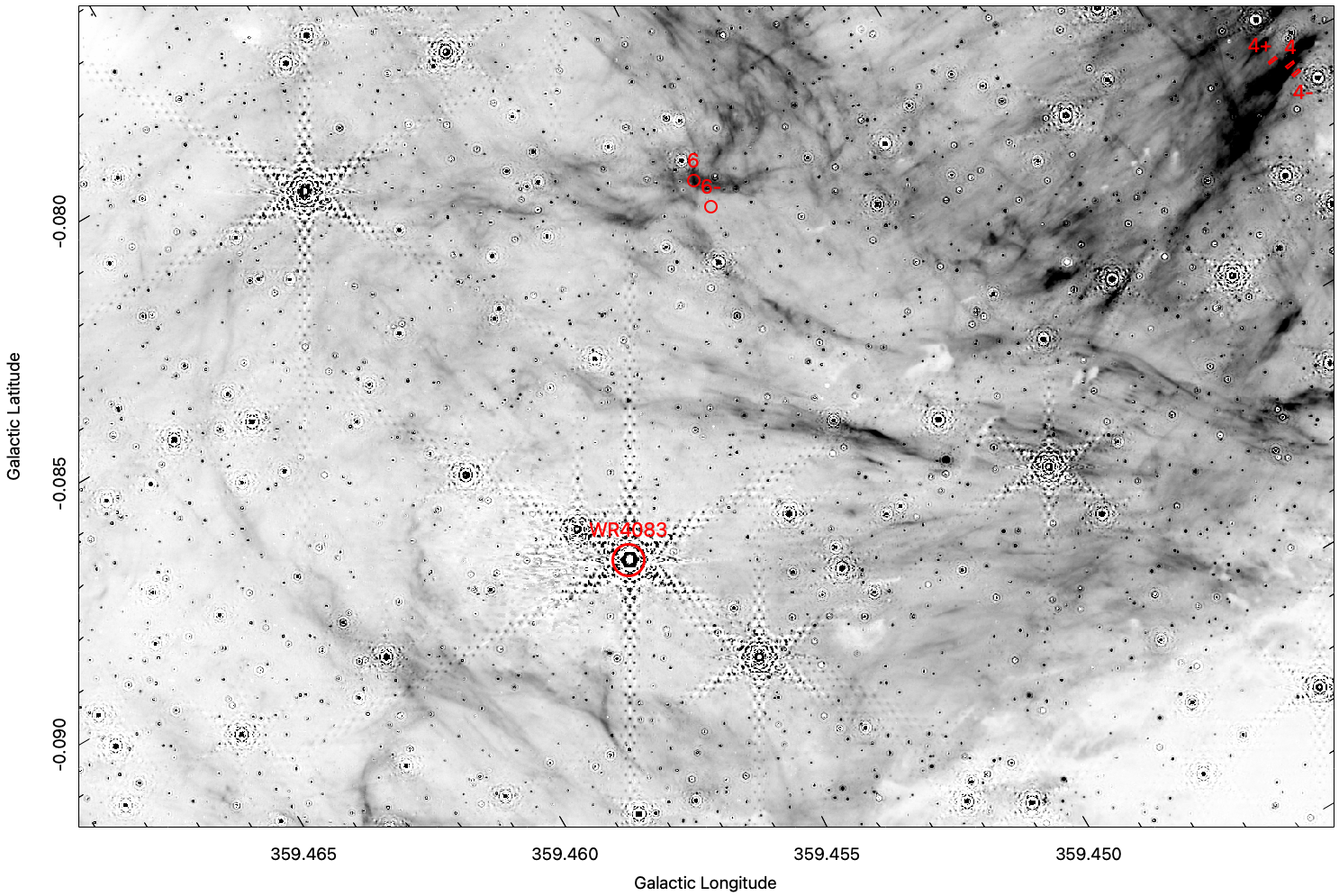}
\caption{\label{figA3} A closeup view of the Br-$\alpha$
emission showing measurement boxes 4 and 6.  This field 
contains radial filaments possibly related to feedback from the 
Wolf-Rayet star WR 4083.}
\end{figure*}

Figure \ref{figA3} shows the Sgr C molecular cloud core, which contains 4 ultra-compact \Hii\ regions that are bright and slightly extended in \Bra .    Note the filaments pointing away from the cloud core.    Several of these structures,  such as the one pointing roughly towards the right from the 
high extinction region (2:40 on a clock-face) resemble bow shocks and may trace outflow lobes emerging from the opaque cloud to become externally irradiated.  The network of stubby, parallel filaments 
pointing toward 12:30 on a clock-face, and located around longitude 359.435, may also trace outflow lobes.  Alternatively, they may trace magnetically confined plasma being photo-ablated from the molecular cloud.  The absence of bright rims or any direct signs of ionization fronts suggests that such fronts are located on the rear side of the cloud and are hidden by many magnitudes of extinction produced by dust in the cloud. 

\begin{figure*}
\centering
\includegraphics[width=1.0\textwidth]{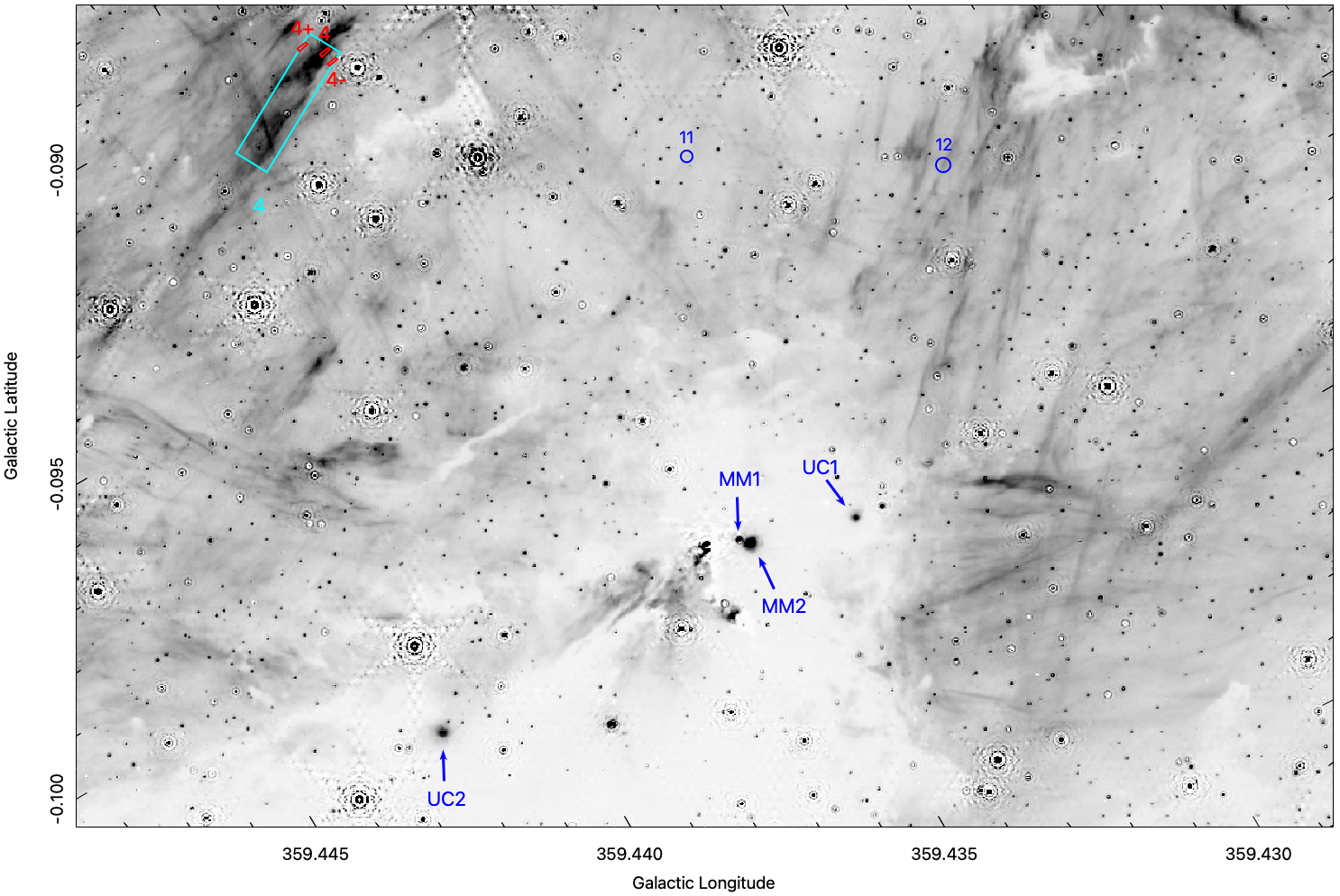}
\caption{\label{figA4} 
A closeup view of the Br-$\alpha$
emission in the region around the Sgr C molecular cloud core, which hosts several 
ultra-compact \Hii\ regions. Measurement box 4 in is the upper left.}
\end{figure*}

Figure \ref{figA4} shows bright filaments radiating away from the Sgr C cloud.  A pair of
wiggly, nearly parallel filaments separated by about 0.02 degrees,  run nearly right to left
above the center of this field.  While the lower structure disappears around longitude 359.460, the
upper filament bends around towards the bottom to connect with a filament located about 0.03 degrees
below the Wolf-Rayet star WR~4083.  This morphology resembles a giant bubble blowing out from the 
cloud surface, located in the lower-right corner of the image.  It may also be energized by the the
bright WR star located below and to the left of the image center.

\begin{figure*}
\centering
\includegraphics[width=1.0\textwidth]{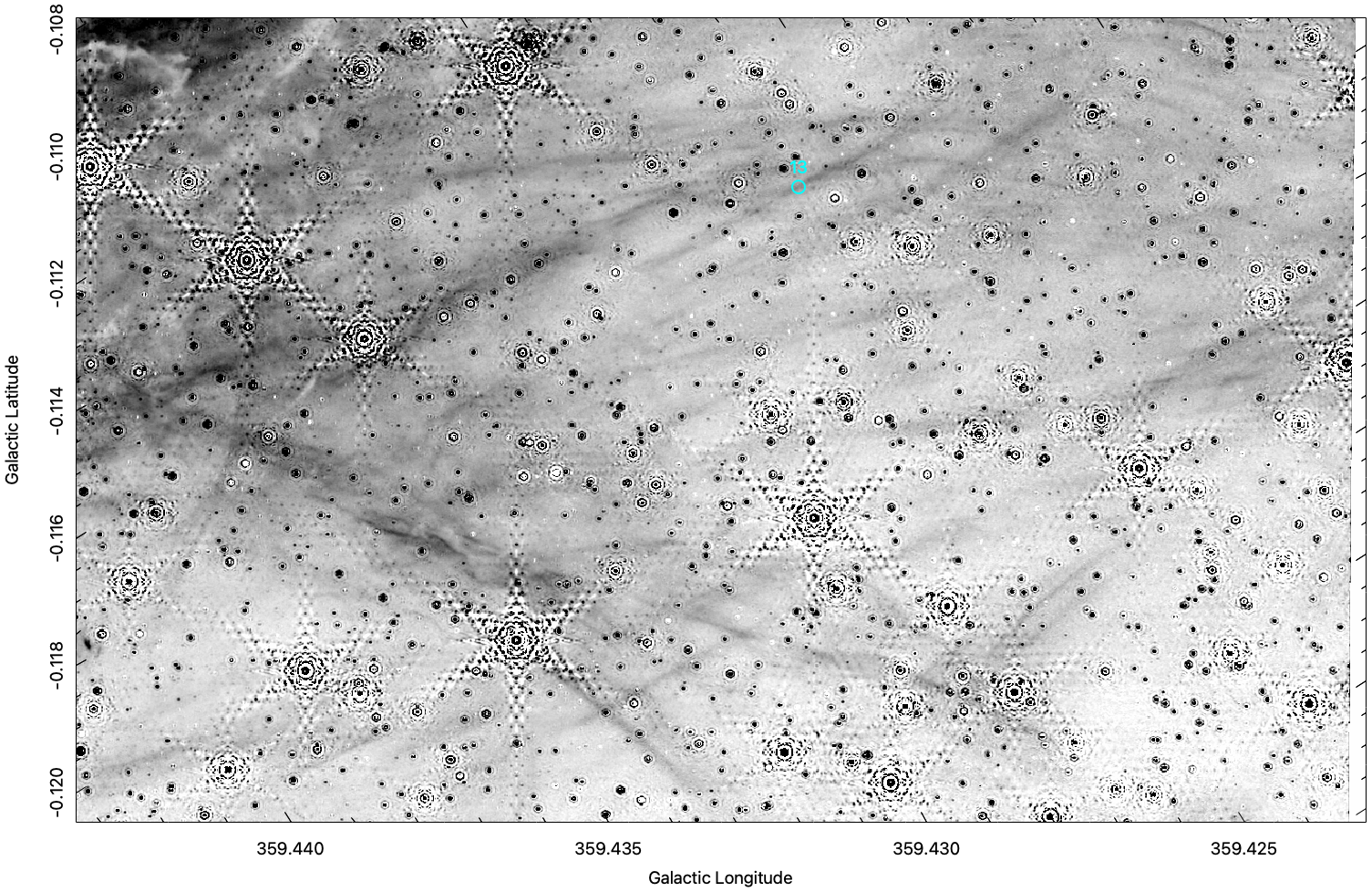}
\caption{\label{figA5} A closeup view of the Br-$\alpha$
emission in the lower-right portion of the JWST field. }
\end{figure*}

Figure \ref{figA5} shows the crosshatching pattern of filaments in the lower-right of the
JWST field.   The filament that drops down towards the lower right is suspected to
trace an ionization front.  The filaments rising towards the right may be magnetically
confined structures.   

\begin{figure*}
\centering
\includegraphics[width=1.0\textwidth]{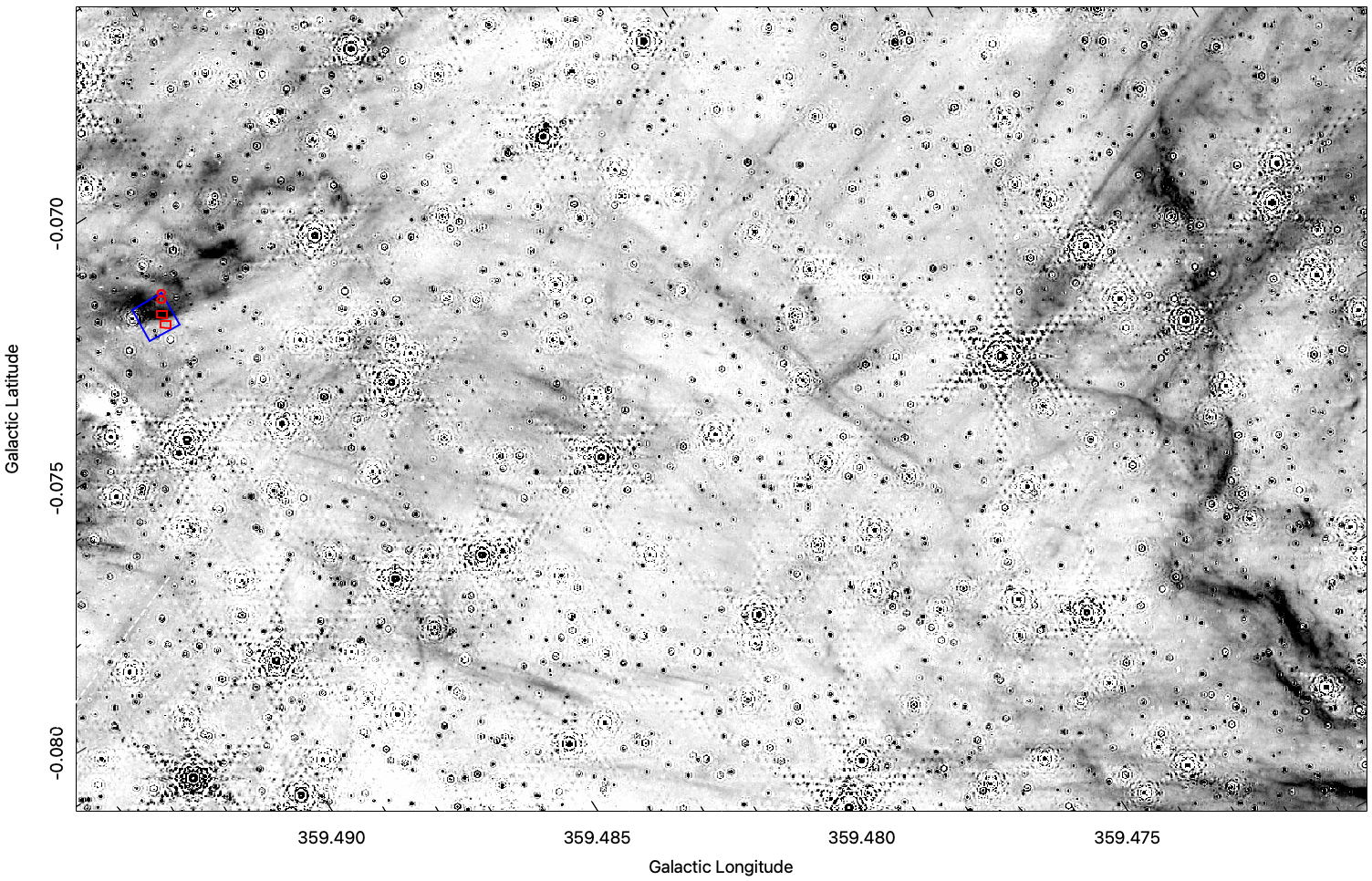}
\caption{\label{figA6} A closeup view of the Br-$\alpha$
emission in the left portion of the JWST field showing the 
ionization front at the high-Galactic-longitude edge of
the Sgr C \Hii\ region (between longitude 359.470 and 
359.475.  It is unclear if the filaments in the left portion of 
the field are excited by massive stars within Sgr C or 
other CMZ OB stars.   Alternatively, they may trace shocks from 
supernova remnants in the CMZ.}
\end{figure*}

Figure \ref{figA6} shows details of the left portion of the JWST field beyond what 
is thought to be the edge of the Sgr C \Hii\ region.   Its unclear where the 
excitation for the \Bra\ emission in this portion of the image comes from.

Table \ref{tab:regions} list the measurements regions shown in the Figures.

\end{document}